\documentclass[twocolumn]{aastex61}

\usepackage{natbib}
\usepackage{graphicx}
\usepackage{times}
\usepackage{apjfonts}
\usepackage{amsmath}
\usepackage{xcolor}
\usepackage{listings}
\usepackage{verbatim}
\usepackage{natbib}
\setcitestyle{citesep={,}}
\providecommand\tabletype{deluxetable*}
\providecommand\tablesize{\scriptsize}

\usepackage{soul}
\usepackage{amsmath}
\usepackage{amssymb}
\usepackage{xspace}
\usepackage{xifthen}
\usepackage{eso-pic}

\newcommand{\nexposures}{38850\xspace}

\newcommand{\areanimagesg}{5224}
\newcommand{\areanimagesr}{5231}
\newcommand{\areanimagesi}{5230}
\newcommand{\areanimagesz}{5234}
\newcommand{\areanimagesy}{5227}
\newcommand{\areanimagesgrizy}{5186}

\newcommand{\astrorepeatg}{42}
\newcommand{\astrorepeatr}{36}
\newcommand{\astrorepeati}{37}
\newcommand{\astrorepeatz}{39}
\newcommand{\astrorepeaty}{56}

\newcommand{\astroabs}{151} 

\newcommand{\photabstatg}{2.6}
\newcommand{\photabstatr}{2.9}
\newcommand{\photabstati}{3.4}
\newcommand{\photabstatz}{2.5}
\newcommand{\photabstaty}{4.5}

\newcommand{\photrepeatg}{7.3}
\newcommand{\photrepeatr}{6.1}
\newcommand{\photrepeati}{5.9}
\newcommand{\photrepeatz}{7.3}
\newcommand{\photrepeaty}{7.8}

\newcommand{\photprecisiong}{5}
\newcommand{\photprecisionr}{4}
\newcommand{\photprecisioni}{4}
\newcommand{\photprecisionz}{5}
\newcommand{\photprecisiony}{5}

\newcommand{\photgaia}{6.6}

\newcommand{\maglimmaxg}{24.32}
\newcommand{\maglimmaxr}{23.89}
\newcommand{\maglimmaxi}{23.41}
\newcommand{\maglimmaxz}{22.09}
\newcommand{\maglimmaxy}{21.40}
\newcommand{\maglimmangg}{24.282}
\newcommand{\maglimmangr}{23.952}
\newcommand{\maglimmangi}{23.335}
\newcommand{\maglimmangz}{22.628}
\newcommand{\maglimmangy}{21.383}
\newcommand{\maglimsnrg}{23.52}
\newcommand{\maglimsnrr}{23.10}
\newcommand{\maglimsnri}{22.51}
\newcommand{\maglimsnrz}{21.81}
\newcommand{\maglimsnry}{20.61}
\newcommand{\maglimsnraperg}{24.33}
\newcommand{\maglimsnraperr}{24.08}
\newcommand{\maglimsnraperi}{23.44}
\newcommand{\maglimsnraperz}{22.69}
\newcommand{\maglimsnrapery}{21.44}

\newcommand{\magcompleteg}{23.72}
\newcommand{\magcompleter}{23.35}
\newcommand{\magcompletei}{22.88}
\newcommand{\magcompletez}{22.25}
\newcommand{\magcompletey}{\ldots}

\newcommand{\medfwhmg}{1.12} 
\newcommand{\medfwhmr}{0.96} 
\newcommand{\medfwhmi}{0.88} 
\newcommand{\medfwhmz}{0.84} 
\newcommand{\medfwhmy}{0.90} 

\newcommand{\medskybrightg}{22.01}
\newcommand{\medskybrightr}{21.15}
\newcommand{\medskybrighti}{19.89}
\newcommand{\medskybrightz}{18.72}
\newcommand{\medskybrighty}{17.96}
\newcommand{\medimagenoiseg}{25.25}
\newcommand{\medimagenoiser}{24.94}
\newcommand{\medimagenoisei}{24.31}
\newcommand{\medimagenoisez}{23.58}
\newcommand{\medimagenoisey}{22.28}

\newcommand{\starefficiency}{90} 
\newcommand{\starcontamination}{3}
\newcommand{\galefficiency}{99}
\newcommand{\galcontamination}{3}

\newcommand{\ie}{i.e.\xspace}
\newcommand{\eg}{e.g.\xspace}

\newcommand{\NEW}[1]{{#1}}
\newcommand{\FIXME}[1]{{\bf \textcolor{red}{#1}}}
\newcommand{\CHECK}[1]{{\textcolor{orange}{#1}}}
\newcommand{\COMMENT}[3]{}

\newcommand{\reportnum}[2]{
  \AddToShipoutPictureBG*{%
    \AtPageUpperLeft{%
      \hspace{0.75\paperwidth}%
      \raisebox{#1\baselineskip}{%
        \makebox[0pt][l]{\textnormal{#2}}
  }}}%
}

\mathchardef\mhyphen="2D

\newcommand{\roughly}{\ensuremath{ {\sim}\,} }

\newlength{\dhatheight}

\newcommand{\code}[1]{\texttt{#1}\xspace}

\newcommand{\unit}[1]{\ensuremath{\mathrm{\,#1}}\xspace}

\newcommand{\GB}{\unit{GB}}
\newcommand{\TB}{\unit{TB}}
\newcommand{\degree}{\ensuremath{{}^{\circ}}\xspace}

\newcommand{\mas}{\unit{mas}}

\newcommand{\asec}{\unit{arcsec}}
\newcommand{\angstrom}{\ensuremath{\,\text{\AA}}\xspace}

\newcommand{\second}{\unit{s}}

\newcommand{\magn}{\unit{mag}}
\newcommand{\mmag}{\unit{mmag}}

\newcommand{\secref}[1]{Section~\ref{sec:#1}}
\newcommand{\appref}[1]{Appendix~\ref{app:#1}}
\newcommand{\tabref}[1]{Table~\ref{tab:#1}}

\newcommand{\figref}[1]{Figure~\ref{fig:#1}}

\newcommand{\bandvar}[2][]{%
  \ifthenelse{\isempty{#1}}{\var{#2}}{\var{#2\_#1}}%
}
\newcommand{\spreadmodel}[1][]{\bandvar[#1]{spread\_model}}

\newcommand{\wavgspreadmodel}[1][]{\bandvar[#1]{wavg\_spread\_model}}
\newcommand{\classstar}[1][]{\bandvar[#1]{class\_star}}
\newcommand{\magauto}[1][]{\bandvar[#1]{mag\_auto}}
\newcommand{\magaper}[1][]{\bandvar[#1]{mag\_aper}}
\newcommand{\magpsf}[1][]{\bandvar[#1]{mag\_psf}}

\newcommand{\imaflagsiso}[1][]{\bandvar[#1]{imaflags\_iso}}
\newcommand{\niter}[1][]{\bandvar[#1]{niter\_model}}

\newcommand{\drmain}[1][]{\bandvar[#1]{dr1\_main}}
\newcommand{\drmagnitude}[1][]{\bandvar[#1]{dr1\_magnitude}}
\newcommand{\drflux}[1][]{\bandvar[#1]{dr1\_flux}}
\newcommand{\drtileinfo}[1][]{\bandvar[#1]{dr1\_tile\_info}}

\newcommand{\SExtractor}{\code{SExtractor}}
\newcommand{\sextractor}{\SExtractor}

\newcommand{\HEALPix}{\code{HEALPix}}
\newcommand{\healpix}{\HEALPix}
\newcommand{\nside}{\code{nside}}

\newcommand{\mangle}{\code{mangle}}

\newcommand{\var}[1]{\ensuremath{\texttt{\MakeUppercase{#1}}}\xspace}

\providecommand\physrep{\ref@jnl{Phys.~Rep.}}%
\providecommand\apjs{\ref@jnl{ApJS}}%
\providecommand{\jcap}{\ref@jnl{JCAP}}%

\newcommand{\KB}[1]{\COMMENT{KB}{orange}{#1}}
\newcommand{\NS}[1]{\COMMENT{NS}{green}{#1}}

\newcommand{\ADW}[1]{\COMMENT{ADW}{magenta}{#1}}

\shorttitle{DES Data Release 1}
\shortauthors{The DES Collaboration}

\reportnum{-9.0}{\footnotesize DES-2017-0315}
\reportnum{-9.7}{\footnotesize FERMILAB-PUB-17-603-AE-E}

\begin{document}

\title{The Dark Energy Survey Data Release 1}


\AuthorCallLimit=300
\author{T.~M.~C.~Abbott}
\affiliation{Cerro Tololo Inter-American Observatory, National Optical Astronomy Observatory, Casilla 603, La Serena, Chile}
\author{F.~B.~Abdalla}
\affiliation{Department of Physics \& Astronomy, University College London, Gower Street, London, WC1E 6BT, UK}
\affiliation{Department of Physics and Electronics, Rhodes University, PO Box 94, Grahamstown, 6140, South Africa}
\author{S.~Allam}
\affiliation{Fermi National Accelerator Laboratory, P. O. Box 500, Batavia, IL 60510, USA}
\author{A.~Amara}
\affiliation{Department of Physics, ETH Zurich, Wolfgang-Pauli-Strasse 16, CH-8093 Zurich, Switzerland}
\author{J.~Annis}
\affiliation{Fermi National Accelerator Laboratory, P. O. Box 500, Batavia, IL 60510, USA}
\author{J.~Asorey}
\affiliation{Centre for Astrophysics \& Supercomputing, Swinburne University of Technology, Victoria 3122, Australia}
\affiliation{School of Mathematics and Physics, University of Queensland,  Brisbane, QLD 4072, Australia}
\affiliation{ARC Centre of Excellence for All-sky Astrophysics (CAASTRO)}
\author{S.~Avila}
\affiliation{Institute of Cosmology \& Gravitation, University of Portsmouth, Portsmouth, PO1 3FX, UK}
\affiliation{Instituto de Fisica Teorica UAM/CSIC, Universidad Autonoma de Madrid, 28049 Madrid, Spain}
\author{O.~Ballester}
\affiliation{Institut de F\'{\i}sica d'Altes Energies (IFAE), The Barcelona Institute of Science and Technology, Campus UAB, 08193 Bellaterra (Barcelona) Spain}
\author{M.~Banerji}
\affiliation{Kavli Institute for Cosmology, University of Cambridge, Madingley Road, Cambridge CB3 0HA, UK}
\affiliation{Institute of Astronomy, University of Cambridge, Madingley Road, Cambridge CB3 0HA, UK}
\author{W.~Barkhouse}
\affiliation{University of North Dakota, Department of Physics and Astrophysics, Witmer Hall, Grand Forks, ND 58202, USA}
% ...

\collaboration{(DES Collaboration)}

\author{S.~Juneau}
\affiliation{National  Optical  Astronomy  Observatory,  950  North  CherryAve, Tucson, AZ 8571, USA}
\author{M.~Fitzpatrick}
\affiliation{National  Optical  Astronomy  Observatory,  950  North  CherryAve, Tucson, AZ 8571, USA}
\author{R.~Nikutta}
\affiliation{National  Optical  Astronomy  Observatory,  950  North  CherryAve, Tucson, AZ 8571, USA}
\author{D.~Nidever}
\affiliation{National  Optical  Astronomy  Observatory,  950  North  CherryAve, Tucson, AZ 8571, USA}
\affiliation{Department of Physics, Montana State University, P.O. Box 173840, Bozeman, MT 59717-3840}
\author{K.~Olsen}
\affiliation{National  Optical  Astronomy  Observatory,  950  North  CherryAve, Tucson, AZ 8571, USA}
\author{A.~Scott}
\affiliation{National  Optical  Astronomy  Observatory,  950  North  CherryAve, Tucson, AZ 8571, USA}
\collaboration{(NOAO Data Lab)}

\correspondingauthor{Matias Carrasco Kind; Keith Bechtol; Ignacio Sevilla}
\email{mcarras2@illinois.edu; kbechtol@lsst.org; nsevilla@gmail.com}


\begin{abstract}

We describe the first public data release of the Dark Energy Survey, DES DR1, consisting of reduced single-epoch images, coadded images, coadded source catalogs, and associated products and services assembled over the first three years of DES science operations.
DES DR1 is based on optical/near-infrared imaging from 345 distinct nights (August 2013 to February 2016) by the Dark Energy Camera mounted on the 4-m Blanco telescope at Cerro Tololo Inter-American Observatory in Chile.
We release data from the DES wide-area survey covering  $\roughly 5000 \deg^2$ of the southern Galactic cap in five broad photometric bands, $grizY$. 
DES DR1 has a median delivered point-spread function of $g = \medfwhmg$, $r = \medfwhmr$, $i = \medfwhmi$, $z = \medfwhmz$, and $Y = \medfwhmy$ \asec FWHM, a photometric precision of $<1$\% in all bands, and an astrometric precision of $\astroabs~\mas$. 
The median coadded catalog depth for a 1.95\arcsec diameter aperture at ${\rm S/N} = 10$ is $g = \maglimsnraperg$, $r = \maglimsnraperr$, $i = \maglimsnraperi$, $z = \maglimsnraperz$, and $Y = \maglimsnrapery$ \magn.
DES DR1 includes nearly $400$M distinct astronomical objects detected in $\roughly10{,}000$ coadd tiles of size  $0.534 \deg^2$ produced from $\roughly39{,}000$ individual exposures.
Benchmark galaxy and stellar samples contain $\roughly310{\rm M}$ and $\roughly80{\rm M}$ objects, respectively, following a basic object quality selection.
These data are accessible through a range of interfaces, including query web clients, image cutout servers, jupyter notebooks, and an interactive coadd image visualization tool.
DES DR1 constitutes the largest photometric data set to date at the achieved depth and photometric precision.

\end{abstract}


%
%
%

\section{Introduction}\label{sec:intro}


Advances in telescope construction, sensor technology, and data processing have allowed us to map the sky with increasing speed and precision, enabling discovery through statistical analysis of astronomical source populations, as well as the detection of rare and/or unexpected objects \citep{2010SPIE.7733E..03T}.
The Dark Energy Survey (DES) is one of several ground-based wide-area optical and near-IR imaging surveys including the Sloan Digital Sky Survey \citep[SDSS,][]{2000AJ....120.1579Y}, the Panoramic Survey Telescope and Rapid Response System 1 \citep[Pan-STARRS1 or PS1,][]{2010SPIE.7733E..0EK}, the Kilo Degree Survey \citep[KiDS, ][]{2013ExA....35...25D}, the Hyper Suprime-Cam Subaru Strategic Program \citep[HSC-SSP;][]{2017arXiv170405858A}, and the future Large Synoptic Survey Telescope \citep[LSST;][]{2008arXiv0805.2366I}.

The instrumental and observational strategies of DES are designed to improve our understanding of cosmic acceleration and the nature of
dark energy using four complementary methods: weak gravitational lensing,
galaxy cluster counts, the large-scale clustering of galaxies (including baryon
acoustic oscillations), and the distances to Type Ia supernovae \citep{2005astro.ph.10346T}.
To achieve these goals, DES conducts two distinct multi-band imaging surveys: a $\roughly 5000 \deg^2$ wide-area survey in the $grizY$ bands and a $\roughly 27 \deg^2$ deep supernova survey observed in the $griz$ bands with a \roughly7-day cadence \citep{Diehl:2014,2015AJ....150..172K}.


DES uses the Dark Energy Camera \citep[DECam;][]{2008arXiv0810.3600H,2015AJ....150..150F},
a 570 Megapixel camera with a $3\deg^2$ field-of-view installed at the prime focus of the Blanco 4-m
telescope at the Cerro Tololo Inter-American Observatory (CTIO) in northern Chile.
Survey observations comprise $\roughly105$ equivalent full nights per year (August through mid-February) including full and half nights.
Each exposure is delivered from CTIO to the National Center for
Supercomputing Applications (NCSA) at the University of Illinois at
Urbana-Champaign for processing generally within minutes of being observed.
At NCSA, the DES Data Management system \citep[DESDM;][]{DESDM,2011arXiv1109.6741S, 2012SPIE.8451E..0DM, 2012ApJ...757...83D} generates a variety of scientific products including single-epoch and coadded images with associated source catalogs of suitable quality to perform precise cosmological measurements \citep[e.g.,][]{Y1KP}.

Raw DES exposures become publicly available one year after acquisition \NEW{from the National Optical Astronomy Observatory (NOAO) Science Archive\footnote{\url{http://archive.noao.edu/}}}, and DES is scheduled to provide two major public releases of processed data.
This first DES Data Release (DR1), described here, encompasses \NEW{data products derived from wide-area survey observations taken in the} first three years of science operations (Y1--Y3\NEW{, from August 2013 to February 2016}). 
A second major data release (DR2) is scheduled for after DES is completed.
In addition to DR1 and DR2, the DES Collaboration prepares incremental internal releases with value-added products and detailed characterizations of survey performance that are designed to support cosmological analyses \citep[e.g., \NEW{Y1} Gold;][]{Y1A1}.
A subset of these products associated with data collected during the DES Science Verification (SV) period (2012 November 1 through 2013 February 22) was released in January 2016.\footnote{\url{https://des.ncsa.illinois.edu/releases/sva1}} \NEW{In September 2018, the value added products from a number of selected DES publications corresponding to Y1 data was released as well.}\footnote{\url{https://des.ncsa.illinois.edu/releases/y1a1}} Additional releases of value-added data products are expected to support \NEW{future} scientific publications.


\begin{\tabletype}{l c c c c c c}
\tablewidth{0pt}
\tabletypesize{\tablesize}
\tablecaption{DES DR1 key numbers and data quality summary. For parameters representing a distribution, the median or mean values are quoted as specified in the main text. \NEW{All magnitudes are in the AB system.} \label{tab:summary}}
\tablehead{
Parameter & \multicolumn{5}{c}{Band} & Reference \\
 & $g$ & $r$ & $i$ & $z$ & $Y$ &
}
\startdata
Number of Exposures in Coadd & 7626 & 7470 & 7470 & 7753 & 8531 & \secref{release} \\
Single-epoch PSF FWHM (\asec) & \medfwhmg & \medfwhmr & \medfwhmi & \medfwhmz & \medfwhmy & \secref{progress} \\
Single-epoch Sky Brightness ($\magn \asec^{-2}$) & \medskybrightg & \medskybrightr & \medskybrighti & \medskybrightz &\medskybrighty & \secref{progress} \\
Single-epoch Effective Image Noise\tablenotemark{a} ($\magn \asec^{-2}$) & \medimagenoiseg & \medimagenoiser & \medimagenoisei & \medimagenoisez &\medimagenoisey & \secref{progress} \\
Sky Coverage (individual bands, deg$^{2}$) & \areanimagesg & \areanimagesr & \areanimagesi & \areanimagesz & \areanimagesy & \secref{progress} \\ 
Sky Coverage ($grizY$ intersection, deg$^{2}$) & \multicolumn{5}{c}{\areanimagesgrizy} & \secref{progress} \\ 
Single-epoch Astrometric Repeatability (total distance, \mas) & \astrorepeatg & \astrorepeatr & \astrorepeati & \astrorepeatz & \astrorepeaty & \secref{astrometry} \\
Coadd Astrometric Precision (total distance, \mas) & \multicolumn{5}{c}{30 (internal); \astroabs (vs. Gaia)} & \secref{astrometry} \\ 
Absolute Photometric Statistical Uncertainty\tablenotemark{b} (mmag) & \photabstatg & \photabstatr & \photabstati & \photabstatz & \photabstaty & \secref{photometry} \\ 
Single-epoch Photometric Repeatability (mmag) & \photrepeatg & \photrepeatr & \photrepeati & \photrepeatz & \photrepeaty & \secref{photometry} \\
Coadd Photometric Precision (mmag) & \photprecisiong & \photprecisionr & \photprecisioni & \photprecisionz & \photprecisiony & \secref{photometry} \\
Coadd Photometric Uniformity vs. Gaia (mmag) & \multicolumn{3}{c}{\photgaia} & \nodata & \nodata & \secref{photometry} \\ 
Single-epoch Magnitude Limit (PSF, ${\rm S/N} = 10$) & 23.57 & 23.34 & 22.78 & 22.10 & 20.69 & \secref{survey} \\
Coadd Magnitude Limit (\var{MAG\_APER\_4}, 1.95 \asec diameter, ${\rm S/N} = 10$) & \maglimsnraperg & \maglimsnraperr & \maglimsnraperi & \maglimsnraperz & \maglimsnrapery & \secref{depth} \\ 
Coadd 95\% Completeness Limit (mag) & \magcompleteg & \magcompleter & \magcompletei & \magcompletez & \nodata & \secref{depth} \\ 
Coadd Spurious Object Rate & \multicolumn{5}{c}{$\lesssim1\%$} & \secref{depth} \\
Coadd Galaxy Selection ($\var{EXTENDED\_COADD} \geq 2$, $\magauto[i] \leq 22.5$) & \multicolumn{5}{c}{Efficiency $>\galefficiency\%$; Contamination $<\galcontamination\%$} & \secref{sgsep} \\
Coadd Stellar Selection ($\var{EXTENDED\_COADD} \leq 1$, $\magauto[i] \leq 22.5$) & \multicolumn{5}{c}{Efficiency $>\starefficiency\%$; Contamination $<\starcontamination\%$} & \secref{sgsep} \\
\enddata
\tablenotetext{a}{Square root of the calibrated image variance, including read noise.}
\tablenotetext{b}{The \textit{Hubble} CalSpec standard star C26202 is used as an absolute reference for the AB system.}
\end{\tabletype}

In this work, we present the content, validation, and data access services for DES DR1.
DR1 is comprised of coadded images and catalogs, as well as calibrated single-epoch images, from the processing of the first three years of DES wide-area survey observations.
Access to DES DR1 data is provided via web interfaces and auxiliary tools, which is made possible through the partnership between NCSA,\footnote{National Center for Supercomputing Applications} LIneA,\footnote{Laborat\'orio Interinstitucional de e-Astronomia} and NOAO\footnote{National Optical Astronomy Observatory}, at the following URL: \url{https://des.ncsa.illinois.edu/releases/dr1}.
In \secref{release}, we briefly describe the DECam instrument and the DES observation strategy for the wide-field survey (the dataset included in this release). \secref{preprocessing} includes an overview of how the raw data were processed by DESDM at NCSA and served as the catalogs and images made available in this release. A basic quality evaluation of these products is presented in \secref{quality}, followed by a description of products as they appear in the DR1 release (\secref{products}). \secref{access} describes the various data access frameworks and tools made available for DR1.  A summary of the release and information on expected future releases is given in \secref{summary}. We direct the reader to \appref{terminology} for definitions of terms and acronyms used throughout the text.

\NEW{Except where noted, all magnitudes quoted in the text are in the AB system \citep{1974ApJS...27...21O}.} \ADW{Is there anywhere that this is noted?}


\section{Data Acquisition}
\label{sec:release}


DR1 is composed of data taken on 345 distinct nights spread over the first three years of DES operations from 2013 August 15 to 2016 February 12.\footnote{DES was scheduled for 319 equivalent full nights, including half-nights, during this period \citep{2016SPIE.9910E..1DD}.}
In this section, we briefly describe the characteristics of the DECam instrument and the DES observation strategy to provide context for DR1.
We point the reader to other DES publications for further details on the technical aspects summarized here \citep[i.e.,][]{2016SPIE.9910E..1DD,Y1A1,DESDM}.


\subsection{DECam}

\begin{figure}
\includegraphics[width=0.5\textwidth]{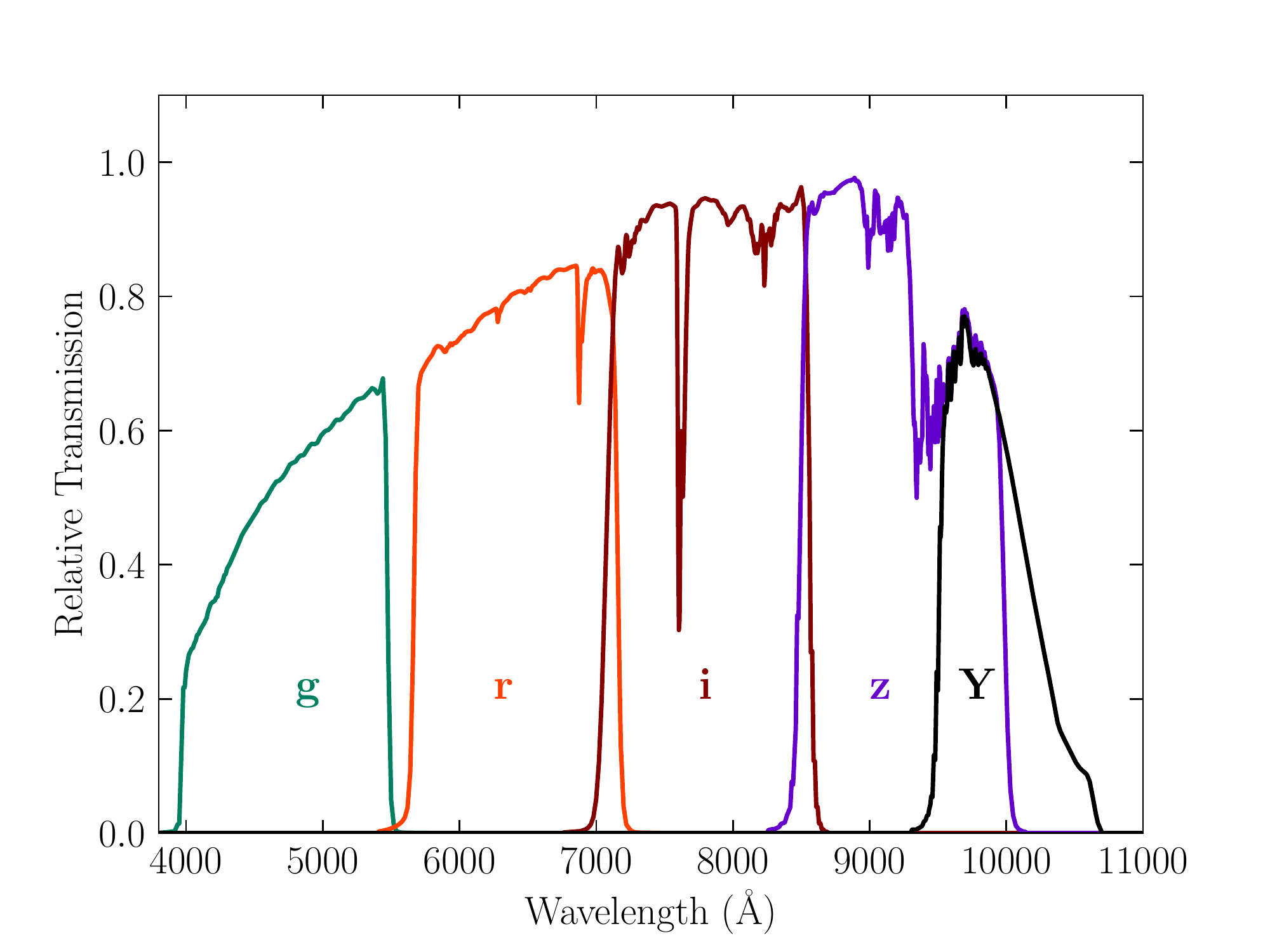}
\caption{DR1 Standard Bandpasses for the DECam $grizY$ filters. The bandpasses represent the total system throughput, including atmospheric transmission (airmass = 1.2) and the average instrumental response across the science CCDs (\secref{bandpass}).}
\label{fig:filters}
\end{figure}

DECam is a wide-field-of-view ($3\deg^2$) mosaic camera containing 62 science CCDs \citep{2015AJ....150..150F}.\footnote{Two and a half DECam CCDs have failed over the course DES operation are only included in DR1 when operating properly \citep{Diehl:2014,DESDM,2015AJ....150..150F}.}
The corrector system and pixel size provide an average plate scale of 0\farcs263 per pixel.
The DES wide-area survey observes five broadband filters, $grizY$ (\figref{filters}), and the standard bandpasses for these filters are included as part of DR1 (\secref{bandpass}).
The DES filters are very similar to their analogously named counterparts from other surveys.

Uniquely, the DES $z$ band has greater sensitivity at longer wavelengths than the SDSS $z$ band and overlaps with the DES $Y$ band.
Additional details, including construction, installation, and a description of DECam subsystems and interfaces are provided in \citet{2015AJ....150..150F}.

\subsection{Survey operations}\label{sec:survey}

\begin{figure*}
\includegraphics[width=\textwidth]{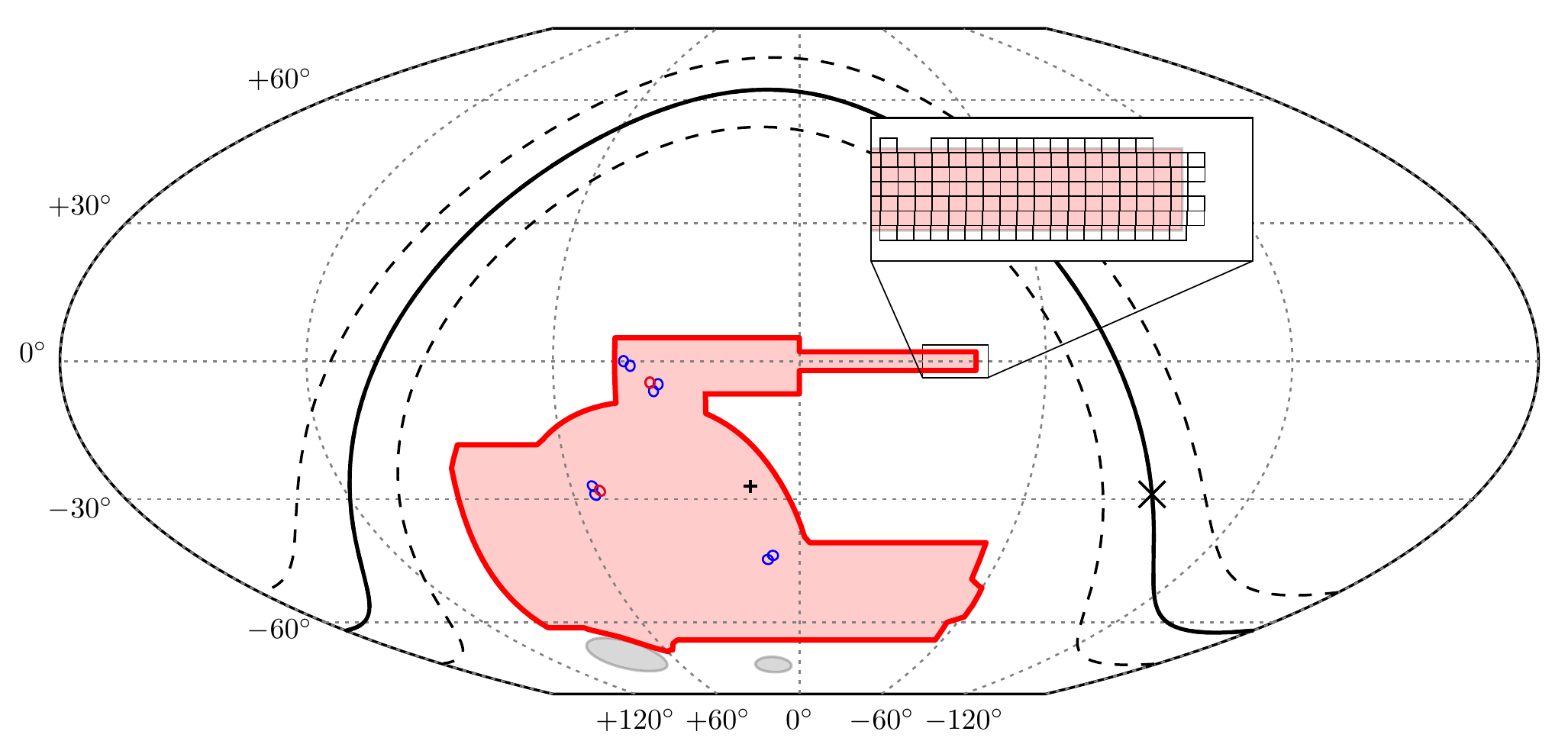}
\caption{A plot of the DES survey area in celestial coordinates. The $\roughly 5000 \deg^2$ wide-area survey footprint is shown in red. The 8 shallow supernova fields are shown as blue circles, and the 2 deep supernova fields are shown as red circles. The Milky Way plane is shown as a solid line, with dashed lines at $b = \pm10 \deg$. The Galactic center (`x') and south Galactic pole (`+') are also marked.
The Large and Small Magellanic Clouds are indicated in gray. 
The inset panel shows an overlay of coadd processing units, coadd tiles, on top of the SDSS Stripe 82 area.
This and the other skymap plots included in this work use the equal-area McBryde-Thomas flat-polar quartic projection.}\label{fig:footprint}
\end{figure*}

The target footprint of the DES wide-area and supernova surveys are shown in \figref{footprint}. All RA, DEC coordinates in this paper refer to the J2000 epoch.  
The wide-area footprint shape was selected to obtain a large overlap with the South Pole Telescope survey \citep{SPTref} and Stripe 82 from SDSS \citep{2009ApJS..182..543A}, and includes a connection region to enhance overall calibration.
Given the cosmological goals of the survey, DES avoids the Galactic plane to minimize stellar foregrounds and extinction from interstellar dust.

The wide-field survey uses exposure times of 90\second for $griz$ and 45\second for $Y$ band, yielding a typical single-epoch PSF depth at ${\rm S/N} = 10$ of $g = 23.57$, $r = 23.34$, $i = 22.78$, $z = 22.10$ and $Y = 20.69$ \citep{DESDM}. 
The completed survey is expected to be roughly one magnitude deeper, through the coaddition of 10 images in each of the bands for a cumulative exposure time of 900~\second in $griz$ and 450~\second in $Y$.\footnote{Beginning in Y4, $Y$-band exposure times were increased to 90\second to reduce overhead while maintaining the same cumulative exposure target.}

Nightly observations are divided between the wide-field and supernova (SN) surveys based on current environmental conditions and the data quality assessments of previous observations.
Realtime optimization of survey strategy is accomplished through the \code{ObsTac} software on the mountain \citep{2014ASPC..485...77N}. 
\code{ObsTac} selects $grizY$ exposures accounting for moon position, sky brightness, current seeing, airmass, hour angle, and other observational characteristics. 
DES exposures are offset by roughly half the focal plane radius on average in successive visits to the same field, or ``hex'', such that objects are observed by different CCDs in each ``tiling''.
This observing strategy minimizes inhomogeneities from the DECam geometry and enhances the relative photometric calibration.
Blanco is an equatorial mount telescope and there is no rotation between dithered and/or repeated exposures.

\begin{figure}
\centerline{
\includegraphics[width=3.8in]{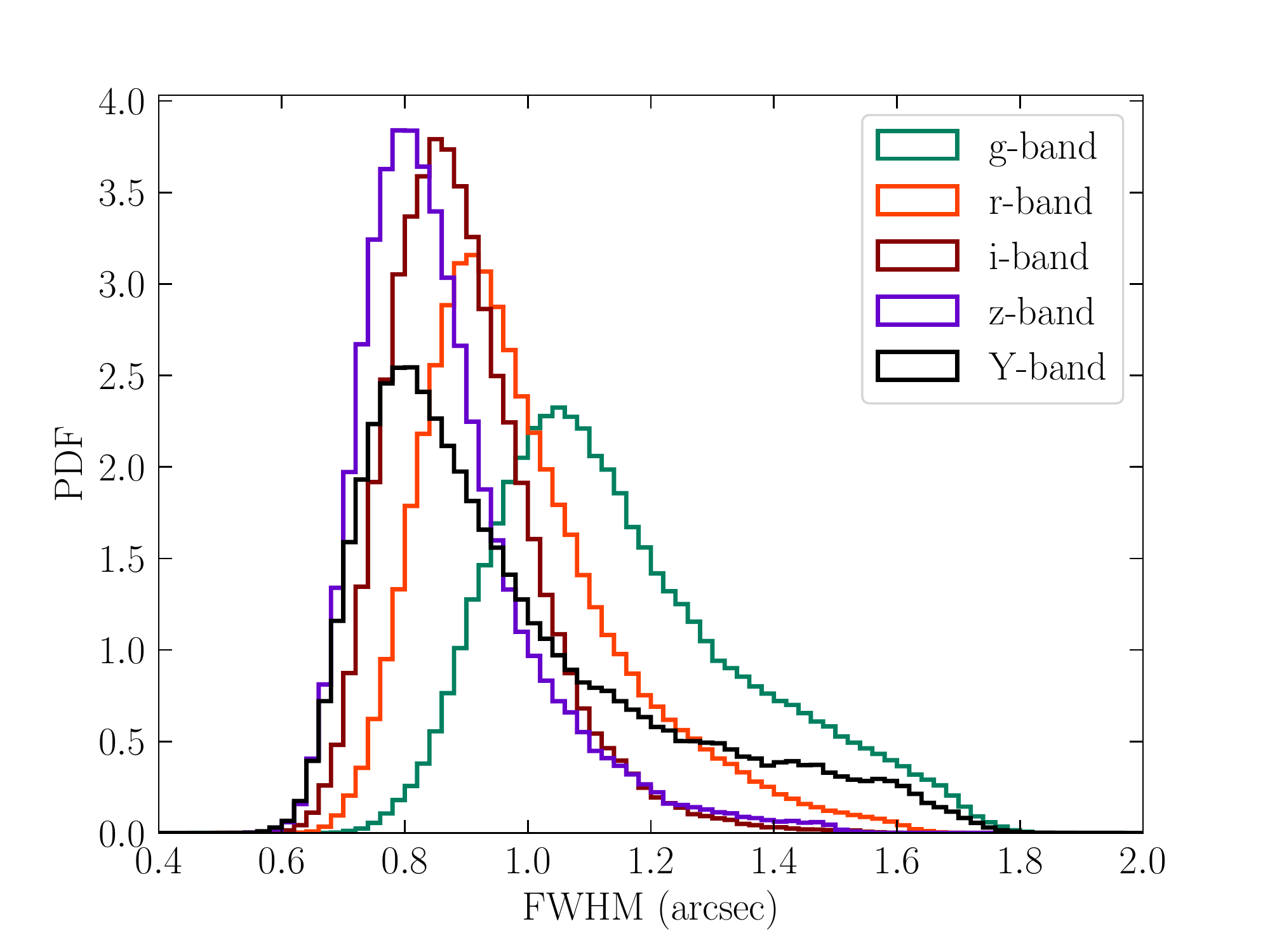}
}
\caption{Normalized histograms showing the distribution of PSF FHWM for single-epoch images that form the DR1 coadd.}\label{fig:fwhm}
\end{figure}

\begin{figure}
\centerline{
\includegraphics[width=3.8in]{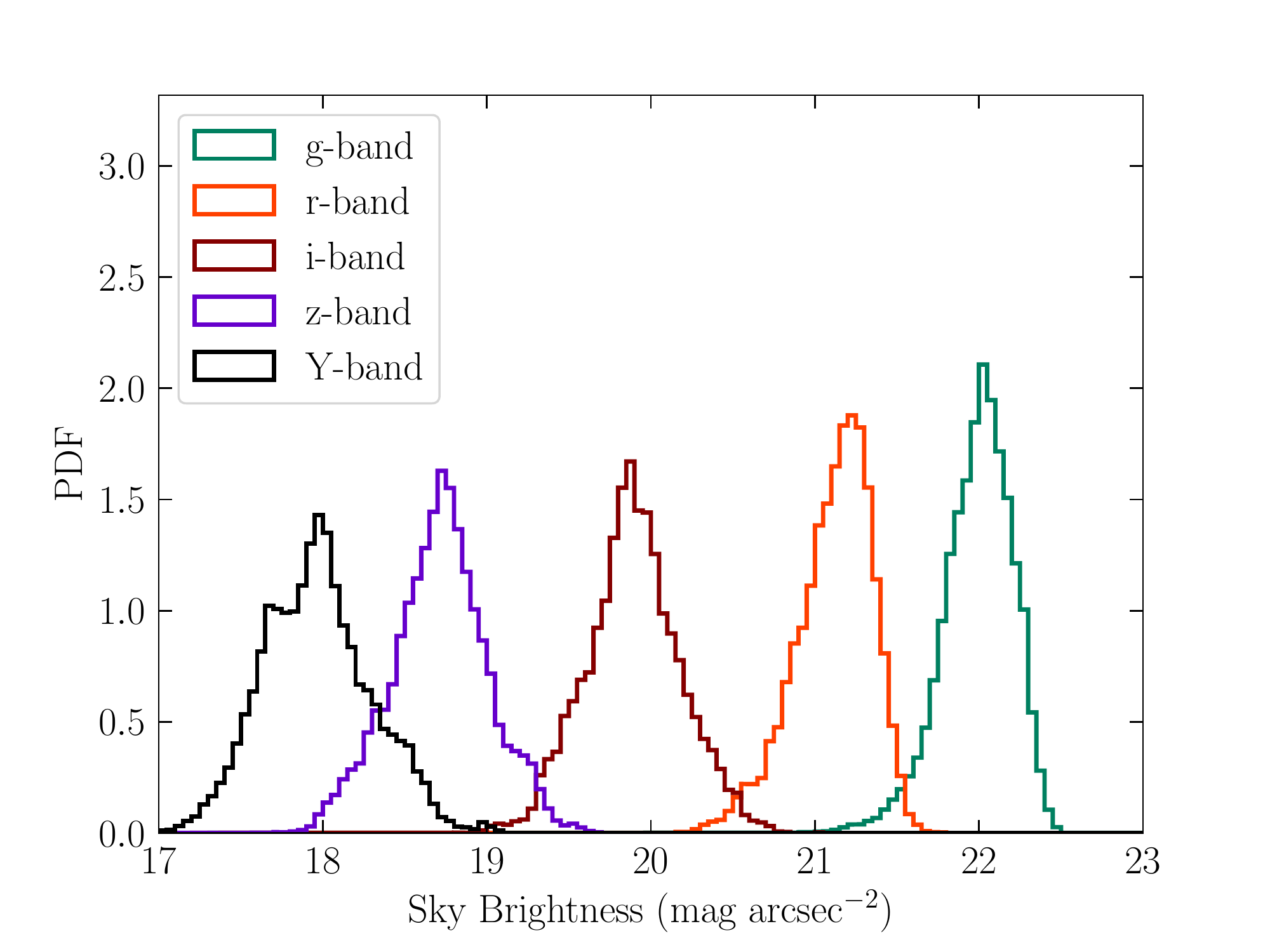}
}
\caption{Normalized histograms showing the distribution of sky brightness for single-epoch images that form the DR1 coadd. \NEW{All magnitudes are given in the AB system.}}\label{fig:sky_brightness}
\end{figure}

A single raw DECam exposure is $\roughly 0.5$ \GB in size (compressed), and DES collects $\roughly 300$ science exposures per night, depending on the season, survey strategy, and on the SN fields schedule.
These data are transferred to NOAO for archiving \citep{2010ASPC..434..260F,2012SPIE.8451E..12H} and to NCSA for further evaluation and processing by the DESDM system; a summary is provided in \secref{preprocessing}.
These raw single-epoch images are made available by NOAO and are accessible as described in \secref{access}.


\subsection{Survey progress through DR1}\label{sec:progress}

Between Y1 and Y3, \nexposures wide-field exposures passed baseline survey quality thresholds \NEW{based on effective exposure time and PSF FWHM} \citep[section 4.7]{DESDM} and are included in coadd processing by DESDM \citep{DESDM}.
The median airmass of DR1 survey-quality exposures was 1.22, with $>99\%$ of exposures taken at airmass $< 1.4$.
Meanwhile, the median delivered seeing (FWHM) was $g = \medfwhmg$, $r = \medfwhmr$, $i = \medfwhmi$, $z = \medfwhmz$, and $Y = \medfwhmy~\asec$ (\figref{fwhm}).
Note that \code{ObsTac} prioritizes observations in the $riz$ bands during periods of good seeing to advance the main science goals of DES (e.g., cosmological constraints from weak gravitational lensing).
\figref{sky_brightness} shows the distribution of sky brightness levels for single-epoch images; the median sky brightness is $g = \medskybrightg$, $r = \medskybrightr$, $i = \medskybrighti$, $z = \medskybrightz$, and $Y = \medskybrighty \magn \asec^{-2}$.
The resulting median single-epoch effective image noise level (square root of the calibrated image variance), including additional contributions from read noise and shot noise of the dome flat, is $g = \medimagenoiseg$, $r = \medimagenoiser$, $i = \medimagenoisei$, $z = \medimagenoisez$, and $Y = \medimagenoisey \magn \asec^{-2}$.

\begin{figure}
\includegraphics[width=0.5\textwidth]{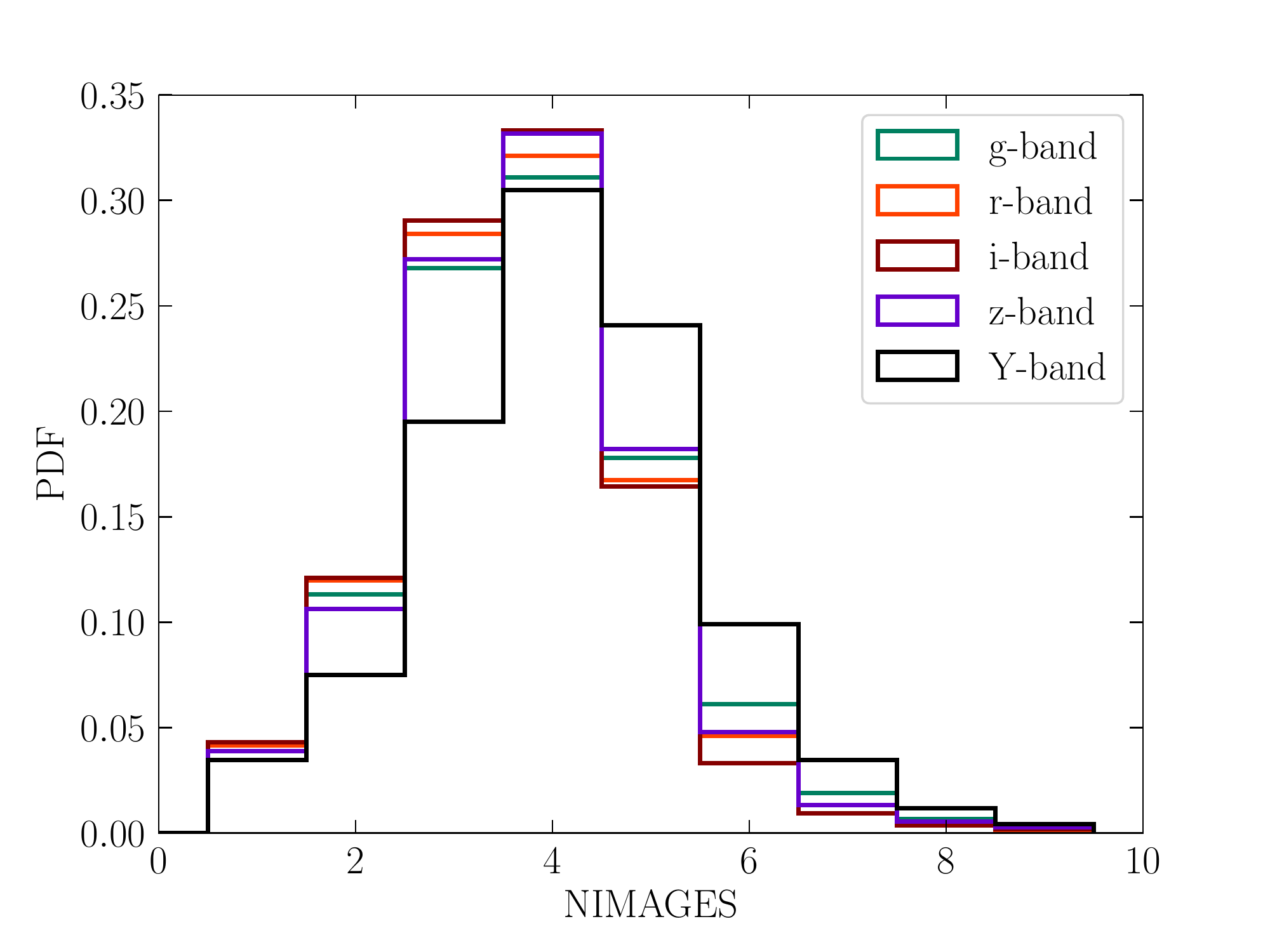}
\caption{Histograms showing the distribution of overlapping images in each of the $grizY$ bands normalized over the DR1 footprint. Most regions of the footprint are covered with 3--5 images.}
\label{fig:nimages_hist}
\end{figure}

\begin{figure}
\includegraphics[width=0.5\textwidth]{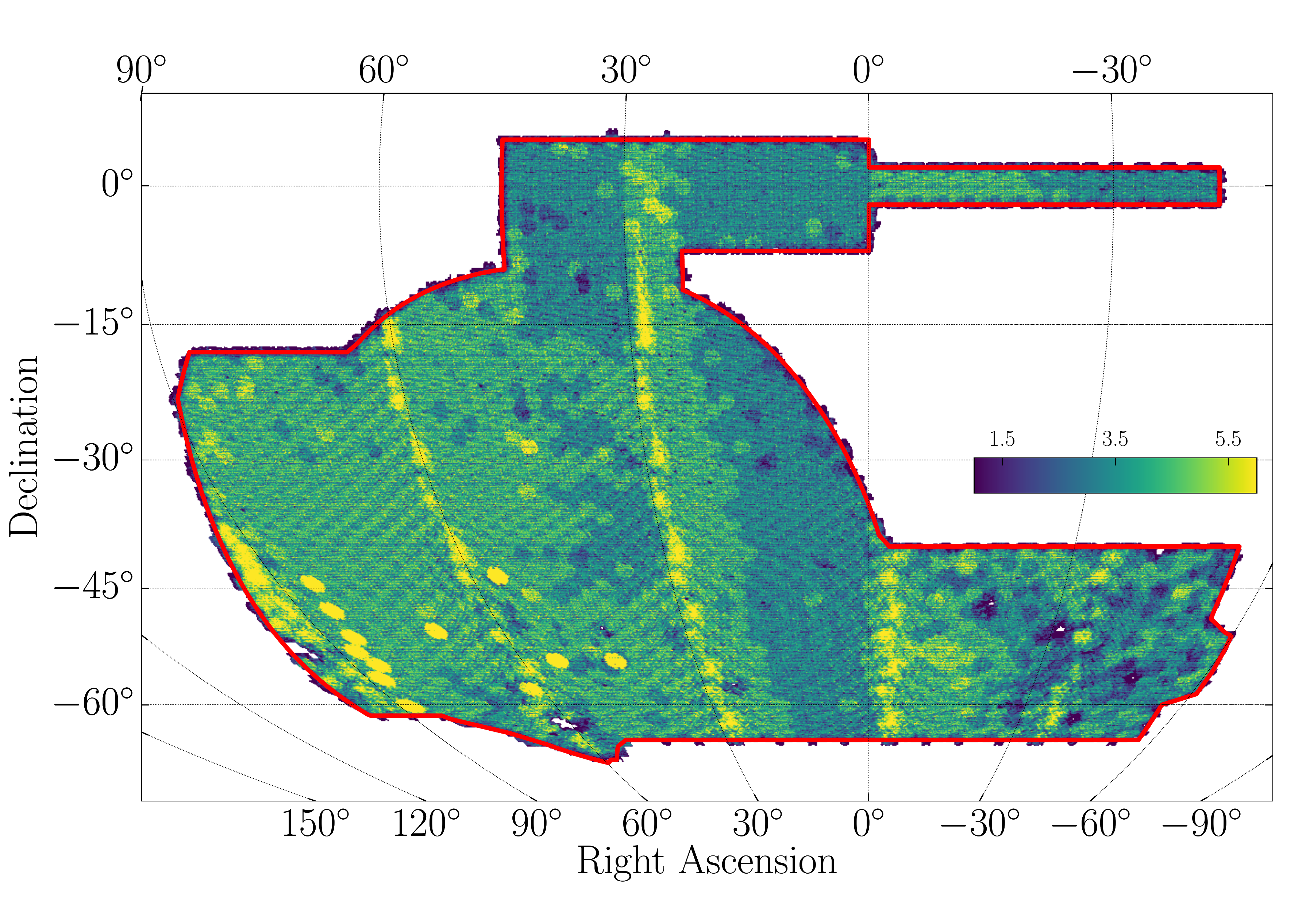}
\caption{Map of the DES footprint showing the number of overlapping $i$-band exposures. Regions of above-average coverage are a consequence of the DES hexagonal layout scheme and can be found at intervals of $\Delta{\rm RA} = 30\degree$. \NEW{Color range units are number of exposures.}}\label{fig:nimages_map}
\end{figure}

\NEW{Each position in the} DES DR1 \NEW{footprint is} typically \NEW{covered by} 3 to 5 overlapping \NEW{DECam} exposures in each of the $grizY$ bands (\figref{nimages_hist}).
As an example, a map for the number of overlapping $i$-band exposures across the footprint is shown in \figref{nimages_map}.

The  total sky coverage of DR1 was estimated using maps of the individual image coverage generated by \mangle \citep{2004MNRAS.349..115H,2008MNRAS.387.1391S} and converted to \healpix \citep{HealpixSoft} maps with spatial resolution comparable to the size of gaps between individual CCDs \citep[\nside = 4096, $\roughly0\farcm86$;][]{Y1A1}.
When requiring at least one exposure in a given band, the areal coverage of each individual band is $g = \areanimagesg$, $r = \areanimagesr$, $i = \areanimagesi$, $z = \areanimagesz$, and $Y = \areanimagesy$ $\deg^{2}$.
When requiring at least one exposure in all of the $grizY$ bands, the DR1 footprint area is 5186 $\deg^{2}$.
These areal coverage numbers do not account for regions that are masked around bright stars or masked due to other imaging artifacts, which decrease the areal coverage by $\roughly 200 \deg^2$. Note also that we are not releasing \mangle products for DR1. Instead, we are providing \healpix indices for all the objects at different resolutions as well as a tabulated \healpix map with \nside = 32 for the footprint. See \secref{access} for more details about these products. 

\section{Data Release Processing}
\label{sec:preprocessing}

We briefly describe the DESDM processing pipeline applied to the DES data to generate the DR1 data products.\footnote{Note that the DESDM pipeline differs from the DECam community pipeline.}  
DR1 is based on the DESDM Y3A2 internal release to the DES Collaboration, referring to the second annual release of data products obtained from the first three seasons of DES science operations and the Science Verification period.
Where possible, our pipelines use the \code{AstrOmatic}\footnote{\url{https://www.astromatic.net/}} suite of tools to perform standard tasks \citep{1996A&AS..117..393B,2002ASPC..281..228B,2006ASPC..351..112B,2010ascl.soft10068B,2011ASPC..442..435B}.
A full description of the pipeline and the underlying image detrending algorithms can be found in \citet{DESDM} and \citet{2017PASP..129k4502B}.  

\subsection{Single-Epoch Processing}
The DES single-epoch processing pipeline (known as ``Final Cut'') removes instrumental signatures to produce reduced, science-ready images \citep{DESDM}.
Final Cut performs: overscan removal, crosstalk correction, non-linearity correction, bias subtraction, gain correction, correction for the brighter-fatter effect \citep{2015JInst..10C5032G}, bad-pixel masking, astrometric matching, flagging of saturated pixels and bleed trails, principal-components background subtraction, secondary flat-field correction, and the masking of cosmic rays and other imaging artifacts.
The resulting images from this pipeline form the products that are released through the NOAO Science Archive\footnote{\url{http://archive.noao.edu/}}.  
Those images are provided in FITS-formatted files and contain extensions for the science data (\code{SCI}), an inverse-variance weight (\code{WGT}), and a mask of bad pixels (\code{MSK}).  
Note that the weight plane is not altered to account for flagged defects; this allows the user to customize the severity of the defects to be removed based on their own analysis needs.  
A summary of the flags available is provided in Table 9 of \citet{DESDM}.

Final Cut also performs PSF model-fitting with \code{PSFEx} \citep{2011ASPC..442..435B} and source detection and measurement with \code{SExtractor} \citep{1996A&AS..117..393B}.
These single epoch data products are not part of DR1.

\subsection{Multi-Epoch (Coadd) Processing}

The multi-epoch pipeline produces coadded images and catalogs of astronomical objects \citep{DESDM}. 
The coadd processing is organized within a tiling scheme that subdivides the sky into square regions with $0\fdg7306$ on a side. 
Coadd images are $10{,}000\times10{,}000$ pixels with a pixel scale is 0\farcs263.  
The choice of images that are tendered as inputs to this
pipeline is based on the data quality assessment that occurred in the Final Cut pipeline and the Forward Global Calibration Method (FGCM) for photometric calibration (\citealt{2018AJ....155...41B}).  
In addition, a `blacklist' of images with severe scattered light, ghosts, or bright transient defects (e.g., comets, meteors, and airplanes) is used to exclude additional images from coadd processing.
The coadded images are re-scaled such that the zeropoint is fixed to $30$ for all filters. 
This makes the conversion between flux and magnitude the same for all bands. 

The multi-epoch pipeline begins by refining the astrometric solution for the image inputs.
This step operates on catalog objects from all input images in all bands simultaneously to 
provide a consistent alignment between images. The relative astrometry within a tile has a 
typical RMS residual of 30 mas or better.  During this process the absolute astrometry is 
tied to 2MASS \citep{2006AJ....131.1163S}\footnote{The Gaia data releases were not yet available when the processing began.}.


The next step prepares the images for coaddition.  First, the World Coordinate System 
\citep[WCS;][]{2002A&A...395.1061G,2002A&A...395.1077C} information for each image is updated 
to reflect the astrometric shifts solved in the previous step.   Then a pair of weight planes
are formed, that set the as-yet unaltered single-epoch weights to zero to remove defects tracked
in the \code{MSK} plane. Both weight planes are formed so that we can separately track spatially 
persistent defects (e.g., saturated stars and bleed trails) and temporary defects (e.g., interpolated 
bad columns, cosmic-rays, satellite trails).  The first weight plane contains all defects, while the second
weight plane contains only the persistent defects.
The \code{AstrOmatic} utility \code{SWarp} \citep{2002ASPC..281..228B}  is then used to form the 
coadd image (\code{SCI}) and weight (\code{WGT}) planes; the former uses the first weight plane 
while the latter uses the second.  A mask plane (\code{MSK}) is formed that carries a value of 0 
for good pixels and 1 for pixels where no good data exist (due to lack of image coverage or persistent defects).


A subsequent execution of \code{SWarp} is then used to form a detection image that is a linear combination
of the $r+i+z$ bands using the `CHI-MEAN' weighting \citep[Appendix B]{Y1A1} and \code{PSFEx} is used to obtain a PSF model for each tile.
Initial catalogs are constructed using \code{SExtractor} in dual image mode where the detection image is used
to form the segmentation map of sources prior to extracting measurements from the individual coadd images.
We caution that the PSF model  is unable to fully account for discontinuities that occur at image boundaries.  
This limits the precision of the measurements of quantities such as \magpsf to no better than a few percent.  In addition, morphological classifiers based on the coadd PSF model, such as
\classstar and \spreadmodel, can have a degraded/varied performance.

To provide a solution that overcomes most of the limitations of the coadd catalog products, the single-epoch catalogs are matched to the coadd detection catalog and weighted averages of the single-epoch \magpsf and
\spreadmodel measurements are made from all unflagged detections of the same object for each band. We add the \var{WAVG\_} prefix to indicate these weighted-average quantities \citep{DESDM}.

These weighted-average measurements are included among the
public data release products.  Longer-term, the DES Collaboration is pursuing improved photometry
through the use of multi-epoch, multi-band, and multi-object fitting that operates on the Final Cut
single-epoch images \citep{Y1A1}.  Those products are maturing but are beyond the scope of the current public data release.
They should become available along with the results from the DES Y3 cosmology analyses.



\section{Data Quality}\label{sec:quality}


\NS{Maybe quality should go after the description of the products (catalogs) to be able to reference the columns}

In this section, we provide a general assessment of the DR1 data quality including astrometric and photometric precision, imaging depth in terms of measurement signal-to-noise and object detection completeness, morphological object classification accuracy, and the identification and removal of likely artifacts.
A summary of data quality metrics is found in \tabref{summary}.

\subsection{Astrometry}\label{sec:astrometry}

\begin{figure*}
\includegraphics[width=\textwidth]{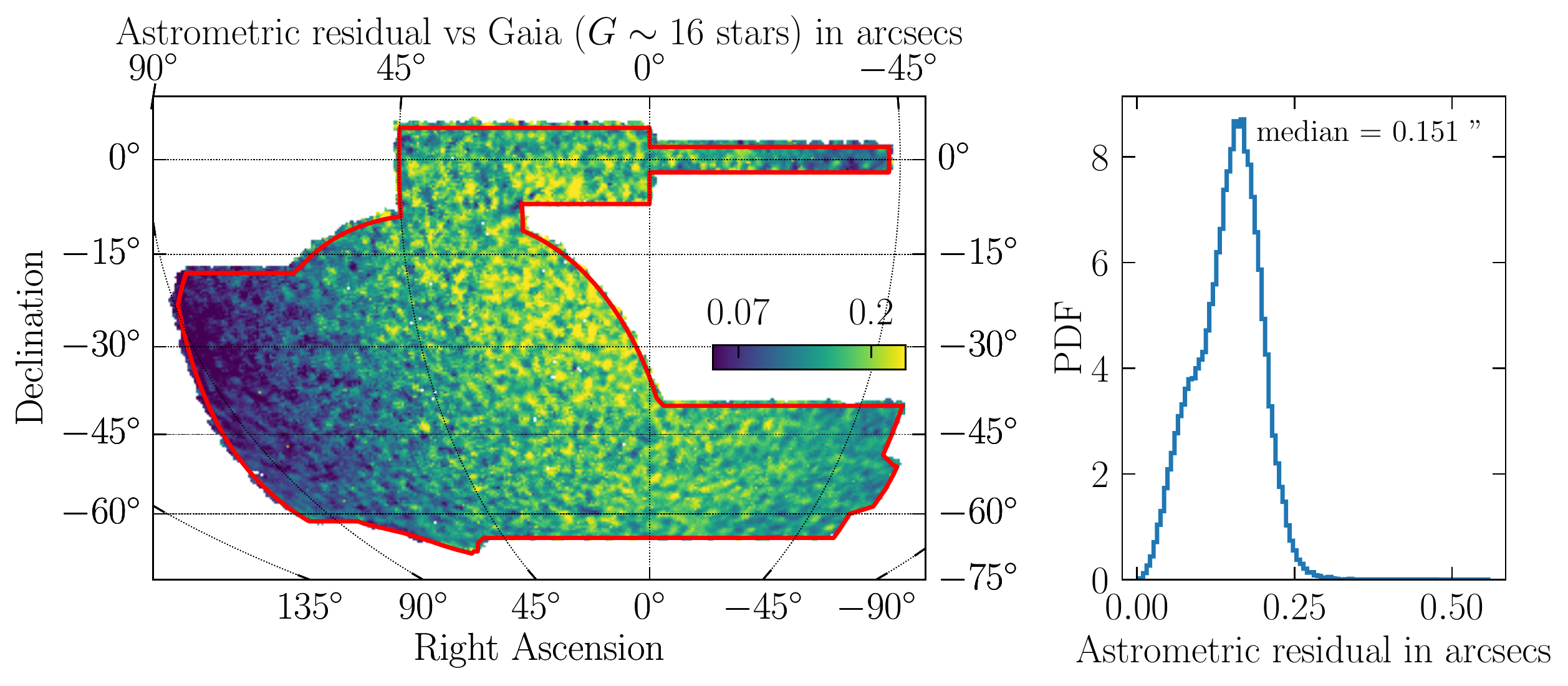}
\caption{Absolute astrometric precision (total distance) measured \NEW{relative to stars} in Gaia DR1 with \NEW{$G$-band magnitude}, $G_{Gaia} \sim 16$ \NEW{\citep{2010A&A...523A..48J}}. \textit{Left}: Mean value of the astrometric discrepancy with respect to Gaia versus sky position computed within \healpix cells of resolution \nside = 256. \textit{Right}: Normalized histogram showing the distribution of astrometric offsets. \NEW{Color range units are in arcsecs.}}\label{fig:astrogaia}
\end{figure*}

The DR1 astrometric solution is derived in two steps using \code{SCAMP} with 2MASS as the reference catalog \citep{DESDM}.
At the single-epoch stage, we find internal astrometric uncertainties of $g = \astrorepeatg$, $r = \astrorepeatr$, $i = \astrorepeati$, $z = \astrorepeatz$, and $Y = \astrorepeaty \mas$, as determined from the median of two-dimensional angular separations between repeated measurements of bright stars from individual exposures.
Following the astrometric refinement step for image coaddition by \code{SCAMP}, the estimated internal astrometric precision for the coadd is $\roughly 30 \mas$ RMS (median over coadd tiles, \NEW{averaging all five bands}).
The absolute astrometric uncertainty of the coadd is evaluated with respect to Gaia DR1 \citep{2016A&A...595A...2G}.
\figref{astrogaia} shows that the two-dimensional astrometric residuals are smallest in regions with larger stellar densities closer to the Galactic plane (for comparison, see \figref{densities}), and the median over the footprint is $\roughly \astroabs \mas$.
Further details of astrometric calibration, including ongoing development using Gaia as a reference catalog, are provided in \citet{2017PASP..129g4503B}.

\subsection{Photometry}\label{sec:photometry}

The relative standard bandpass flux measurements for astronomical sources in DR1 have been calibrated using a forward modeling technique to account for both atmospheric and instrumental components of the total system throughput \citep[FGCM;][]{2018AJ....155...41B}.
The parameters of the model were initially fit from observations taken in photometric conditions to establish a network of calibration stars that spans the full survey footprint. This network was subsequently used to refine the calibration of exposures taken in non-photometric conditions.
Meanwhile, the absolute photometric calibration of DR1 is tied to the spectrophotometric \textit{Hubble} CALSPEC standard star C26202 \citep{2014PASP..126..711B} located in the SN field C3.

Over one hundred repeated measurements of C26202 in a variety of conditions yielded a set of small shifts ($\sim 3 \mmag$) to place the DES photometry on the AB system. 
These shifts have been pre-applied to the DR1 zeropoints.
The statistical uncertainty on these shifts is estimated to be $g = \photabstatg$, $r = \photabstatr$, $i = \photabstati$, $z = \photabstatz$, and $Y = \photabstaty$~\mmag.
Additional sources of systematic uncertainty on the absolute photometric calibration could arise from uncertainty in the level of out-of-band light leakage and uncertainty in the synthetic photometry of C26202.
We are currently undertaking observations and analysis of two additional HST CALSPEC standards to reduce the systematic uncertainty of the AB offsets.

\begin{figure*}
\includegraphics[width=\textwidth]{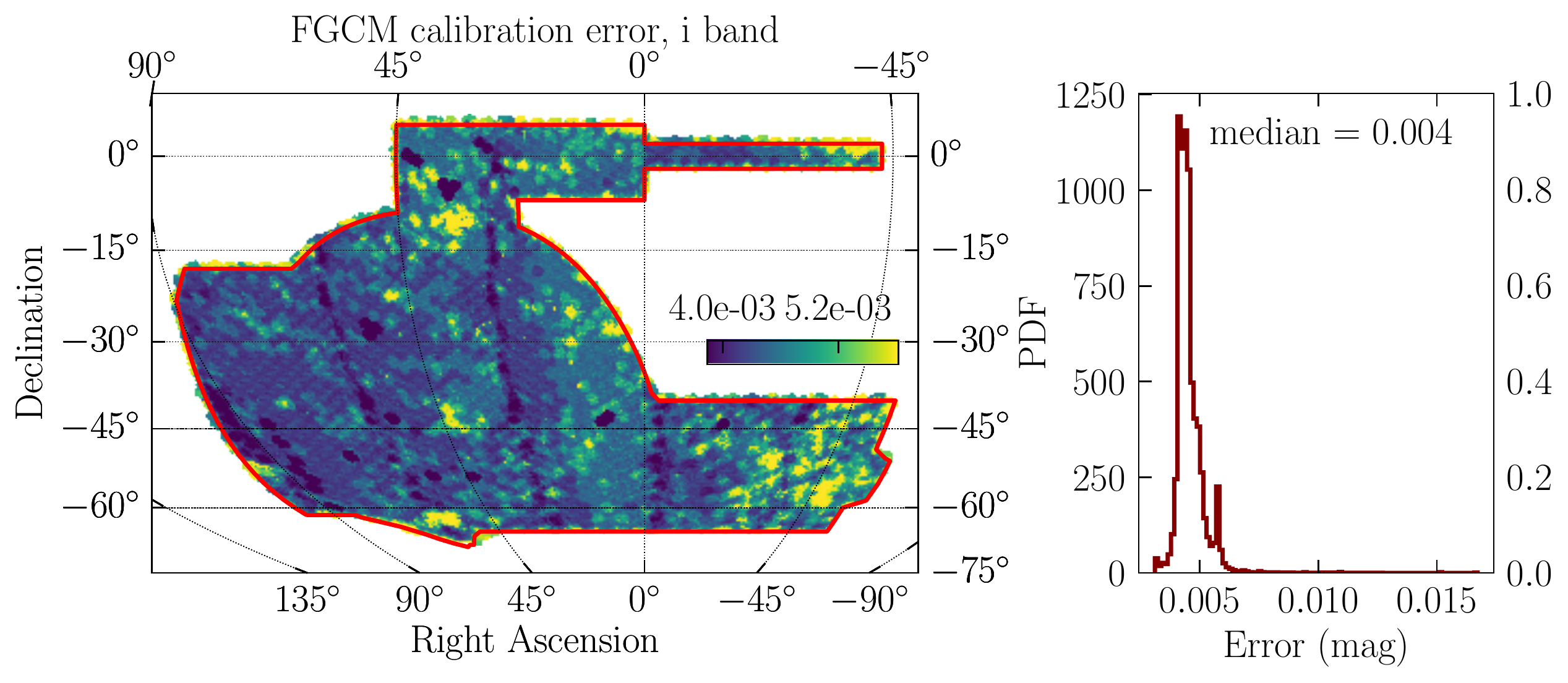}
\caption{Statistical uncertainty of coadd zeropoints in the $i$ band estimated from the FGCM photometric calibration. \textit{Left}: Mean value of the uncertainty vs. sky position, computed within \healpix cells of resolution \nside = 256. \textit{Right}: Normalized histogram showing the distribution of zeropoint uncertainties over the footprint. \NEW{Color range units are in AB magnitudes.}}\label{fig:fgcm_caliberror}
\end{figure*}

\begin{figure*}
\includegraphics[width=\textwidth]{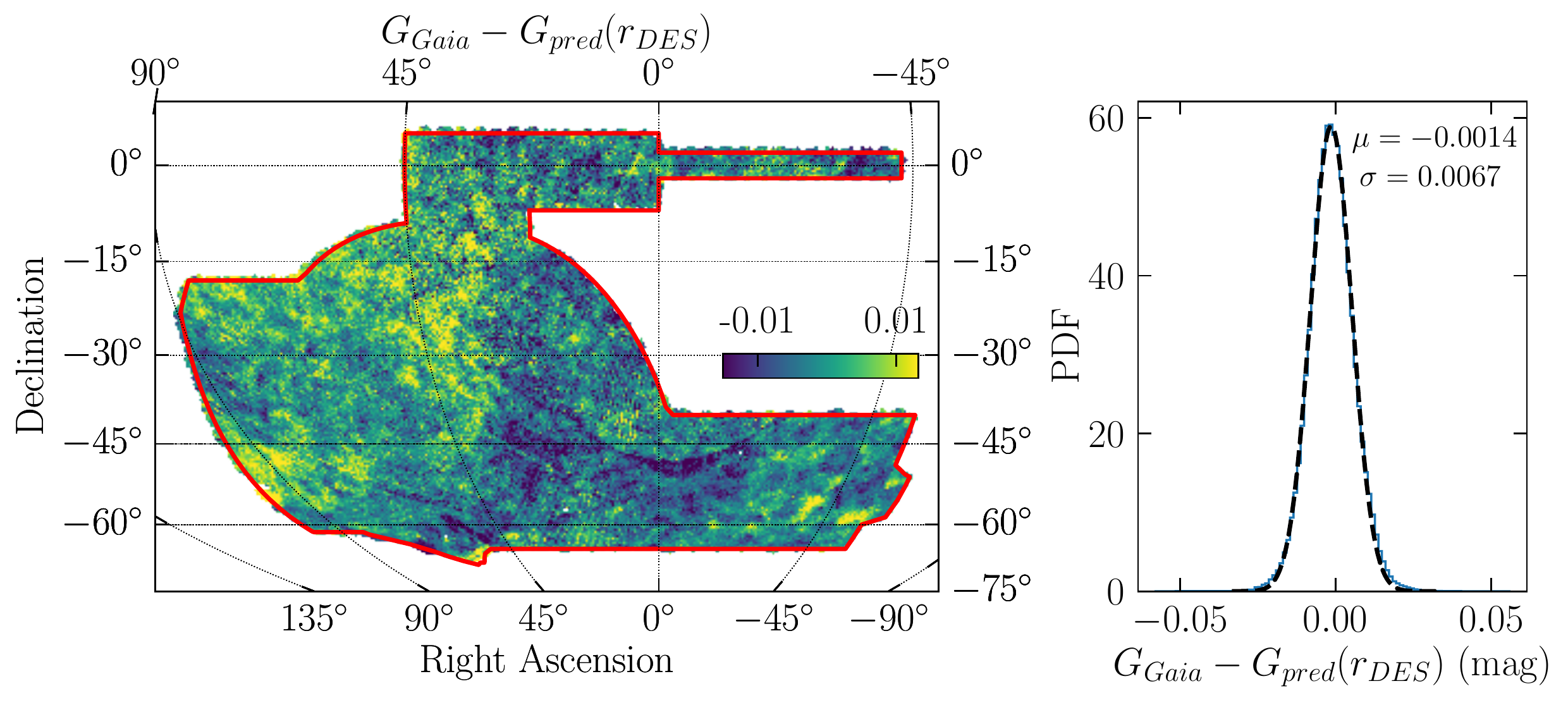}
\caption{The distribution of photometric residuals against Gaia's $G$ band  (\magn units) across the DES footprint is used the evaluate the uniformity of the coadd FGCM calibration. \textit{Left}: Mean value of this residual vs sky position in a \healpix cell of \nside = 256. \textit{Right}: Normalized histogram of photometric residuals over the footprint. \NEW{Color range units are in AB magnitudes.}}\label{fig:photogaia}
\end{figure*}

We reproduce below several of the key results from \citet{2018AJ....155...41B} regarding the relative photometric calibration.
The single-epoch photometric statistical precision (associated with random errors in the FGCM fit parameters) derived from repeated measurements of FGCM calibration stars is $g = \photrepeatg$, $r = \photrepeatr$, $i = \photrepeati$, $z = \photrepeatz$, and $Y = \photrepeaty$~\mmag.
Under the assumption that successive tiled observations of the same fields yield largely independent model fit parameters (as would be expected from the widely spaced observations in DES), we estimate the statistical precision of coadd zeropoints by combining the fit results from overlapping exposures.
The median coadd zeropoint statistical uncertainty is $g = \photprecisiong$, $r = \photprecisionr$, $i = \photprecisioni$, $z = \photprecisionz$, and $Y = \photprecisiony$~\mmag.
\figref{fgcm_caliberror} shows an example distribution for the $i$-band.
As a validation, we compare the photometric uniformity of DES DR1 to the space-based Gaia $G$-band photometry (\figref{photogaia}). Variations in uniformity are found to be \photgaia~\mmag, as estimated from a Gaussian fit to the offset distribution between $G_{Gaia}$ and $G_{\rm pred}(r)$ predicted from the DES $r$ band (\appref{transform}).





To account for extinction from interstellar dust, we include an additional column in the DR1 photometry tables for $E(B-V)$ values from the reddening map of \citet{1998ApJ...500..525S} (SFD98) at the location of each catalog object.
The $E(B-V)$ values were obtained using a linear interpolation of the Zenithal Equal Area projected map distributed by SFD98.
We computed fiducial interstellar extinction coefficients, $R_b$, for each band so that the corrections to the FGCM calibrated source magnitudes are $A_b = E(B-V) R_b$.
Fiducial coefficients are derived using the \citet{1999PASP..111...63F} reddening law with $R_V = 3.1$ and the \citet{2011ApJ...737..103S} calibration adjustment to the original SFD98 reddening map ($N = 0.78$).
Note that the \citet{2011ApJ...737..103S} calibration adjustment is included in our fiducial reddening coefficients; these coefficients are intended to be used directly with $E(B-V)$ values from the original SFD98 reddening map.
We integrate over the DR1 Standard Bandpasses (\secref{bandpass}) considering a fixed source spectrum that is constant in spectral flux density per unit wavelength, $f_{\lambda}$ (erg~cm$^{-2}$~s$^{-1}$~\AA$^{-1}$), and adopted the low-extinction limit.
The latter simplification is appropriate for DES, for which $E(B-V) < 0.1$~mag over $\sim99\%$ of the footprint.
The resulting multiplicative coefficient for each band is $R_g = 3.186$,  $R_r = 2.140$,  $R_i = 1.569$,  $R_z = 1.196$ and  $R_Y = 1.048$.
The DES science team continues to explore systematic uncertainties associated with interstellar extinction modeling.

In general, the photometry columns included in the DR1 database tables are not dereddened by default.
However, the \drmain table includes additional columns for the dereddened versions of \magauto and \var{wavg\_mag\_psf} indicated by a \var{\_DERED} suffix (\appref{appendix_tables}). 

\subsection{Flagged objects}\label{sec:flags}

For coadd objects, if any pixel is masked in all of the contributing exposures for a given band, the \imaflagsiso = 1 flag is set for that band.
This flag is predominantly set for saturated objects and objects with missing data.
We recommend a baseline quality criteria of \imaflagsiso = 0 (in the relevant bands) for most science applications since the majority of flagged objects have unreliable photometry.
The DES DR1 catalog also includes other standard flags \var{FLAGS} provided by the \code{SExtractor} pipeline.
\NEW{A summary of \sextractor \var{FLAGS} bitmask values and warning descriptions is provided in \appref{sextractor_flags}.}

\subsection{Depth}\label{sec:depth}

The effective depth of the DES DR1 wide-field coadd catalog is dependent on the photometric measurement of interest and can be quantified through various approaches. 
Here we derive simple depth estimates from the flux distribution of cataloged objects, the magnitude corresponding to a fixed signal-to-noise threshold (${\rm S/N} = 10$), an aperture estimate from the input imaging, and object detection completeness.
In general, the type of source and surface brightness must also be considered when evaluating the survey depth.

The settings used for the source extraction and deblending steps of the DESDM pipeline allow for efficient detection of objects with ${\rm S/N} \sim 10$ in the $r+i+z$ composite detection image \citep{DESDM}. 
At the bright end, saturation effects start to become important at $r < 16$. 
For even brighter magnitudes, the fraction of saturated objects increases until objects are no longer cataloged due to pixel-level masking applied during processing.
For each of the depth studies considered here, we selected a sample of high-quality DR1 coadd objects using $\var{flags\_[griz]} < 4$ and \var{imaflags\_iso\_[griz]} = 0. 
No star-galaxy selection has been applied to the sample of analyzed objects.
Below we describe these different approaches to estimate the depth of DR1; results from each method are provided in \tabref{depth_estimates}.

\begin{\tabletype}{l c c c c c}
\tablewidth{0pt}
\tabletypesize{\tablesize}
\tablehead{
Method & \multicolumn{5}{c}{Band} \\
 & $g$ & $r$ & $i$ & $z$ & $Y$}
\tablecaption{DES DR1 coadd catalog median depth estimates for the sample of all high-quality objects. \NEW{All magnitudes are given in the AB system.} \label{tab:depth_estimates}}
\startdata
Maximum in number counts (\var{MAG\_AUTO})   & \maglimmaxg  & \maglimmaxr  & \maglimmaxi  & \maglimmaxz  & \maglimmaxy  \\
Measured with ${\rm S/N} = 10$ (\var{MAG\_AUTO}) & \maglimsnrg  & \maglimsnrr  & \maglimsnri  &  \maglimsnrz & \maglimsnry  \\
Measured with ${\rm S/N} = 10$ (\var{MAG\_APER\_4}) & \maglimsnraperg  & \maglimsnraperr  & \maglimsnraperi  &  \maglimsnraperz & \maglimsnrapery  \\
Imaging depth from \mangle (\var{MAG\_APER\_4})     & \maglimmangg & \maglimmangr & \maglimmangi & \maglimmangz & \maglimmangy \\
Detection completeness of 95\% (\var{MAG\_AUTO}) & \magcompleteg & \magcompleter & \magcompletei & \magcompletez & \magcompletey \\
\enddata
\end{\tabletype}

\subsubsection{Flux distribution}

\begin{figure}
\centerline{
\includegraphics[width=0.5\textwidth]{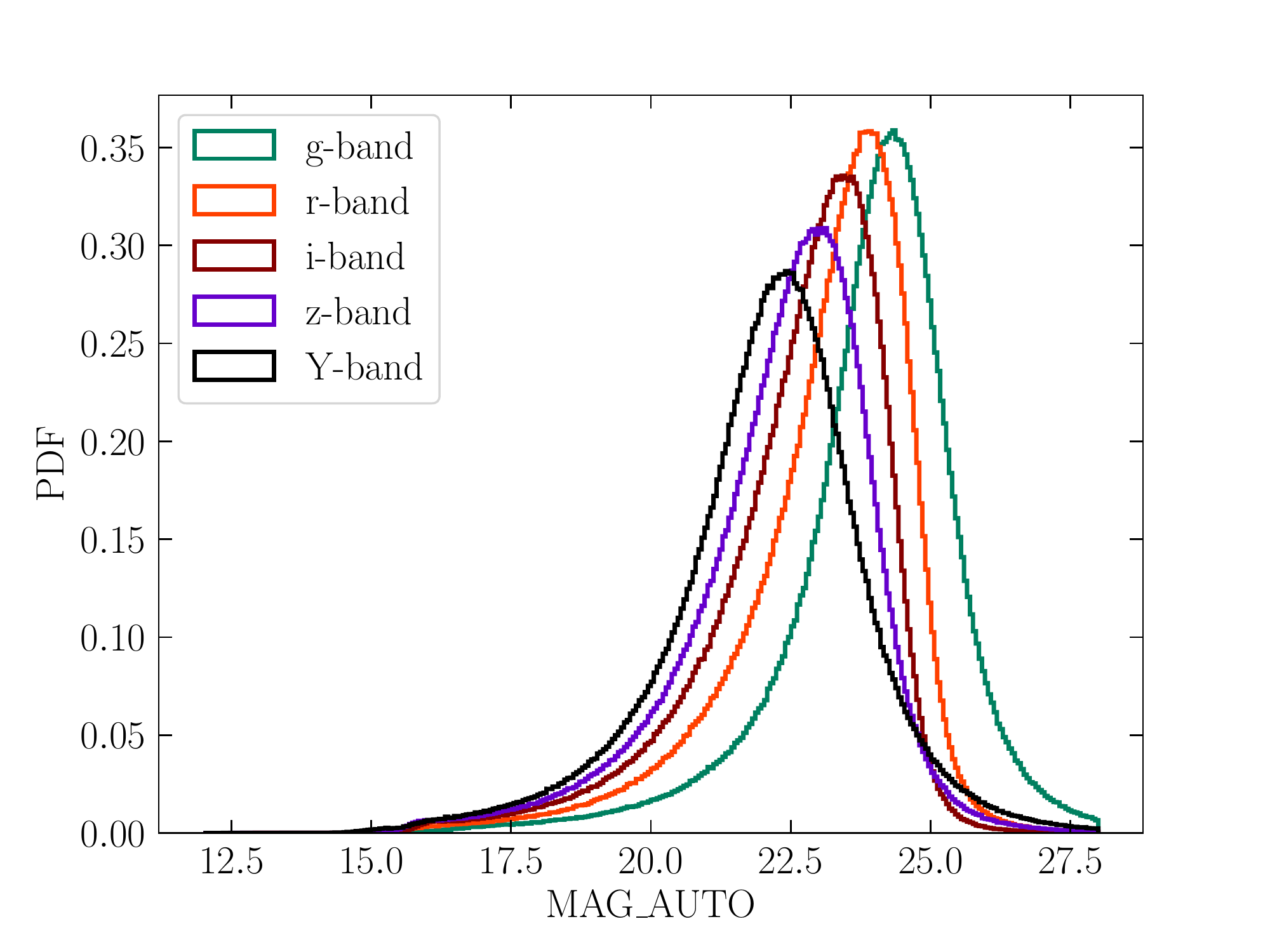}
}
\caption{Normalized histograms of source counts binned by \texttt{SExtractor}'s \magauto quantity showing the flux distribution of detected sources. \NEW{All magnitudes are given in the AB system.}}\label{fig:dr1_magauto}
\end{figure}

The distribution of astronomical sources is weighted towards low-flux sources.
A crude estimate of the detection threshold is given by the mode  of the number counts distribution of sources as a function of magnitude.
\figref{dr1_magauto} shows the number counts of coadd objects as a function of \magauto in all five DES bands.
The mode of the \magauto distribution is $g = \maglimmaxg$, $r = \maglimmaxr$, $i = \maglimmaxi$, $z = \maglimmaxz$, and $Y = \maglimmaxy$.
No restrictions were placed on source morphology (\ie, stars vs.\ galaxies) for this estimate.


\subsubsection{Magnitude limit at fixed signal-to-noise}


The magnitude limit corresponding to a fixed signal-to-noise for a given photometric measurement (e.g., \magauto) can be empirically determined from the distribution of magnitude uncertainties as a function of magnitude \citep{2015arXiv150900870R}.
\code{SExtractor} provides an estimate of the photometric uncertainty through the \var{MAGERR} quantities, which are estimated from the fluctuations of the background around the sources. 
These are related to signal-to-noise, $\delta F/F$, via the differentiation of Pogson's law \citep{1856MNRAS..17...12P}:

\begin{equation}
\delta m = \frac{-2.5}{\ln 10} \frac{\delta F}{F}.
\end{equation}

\begin{figure*}
\includegraphics[width=0.5\textwidth]{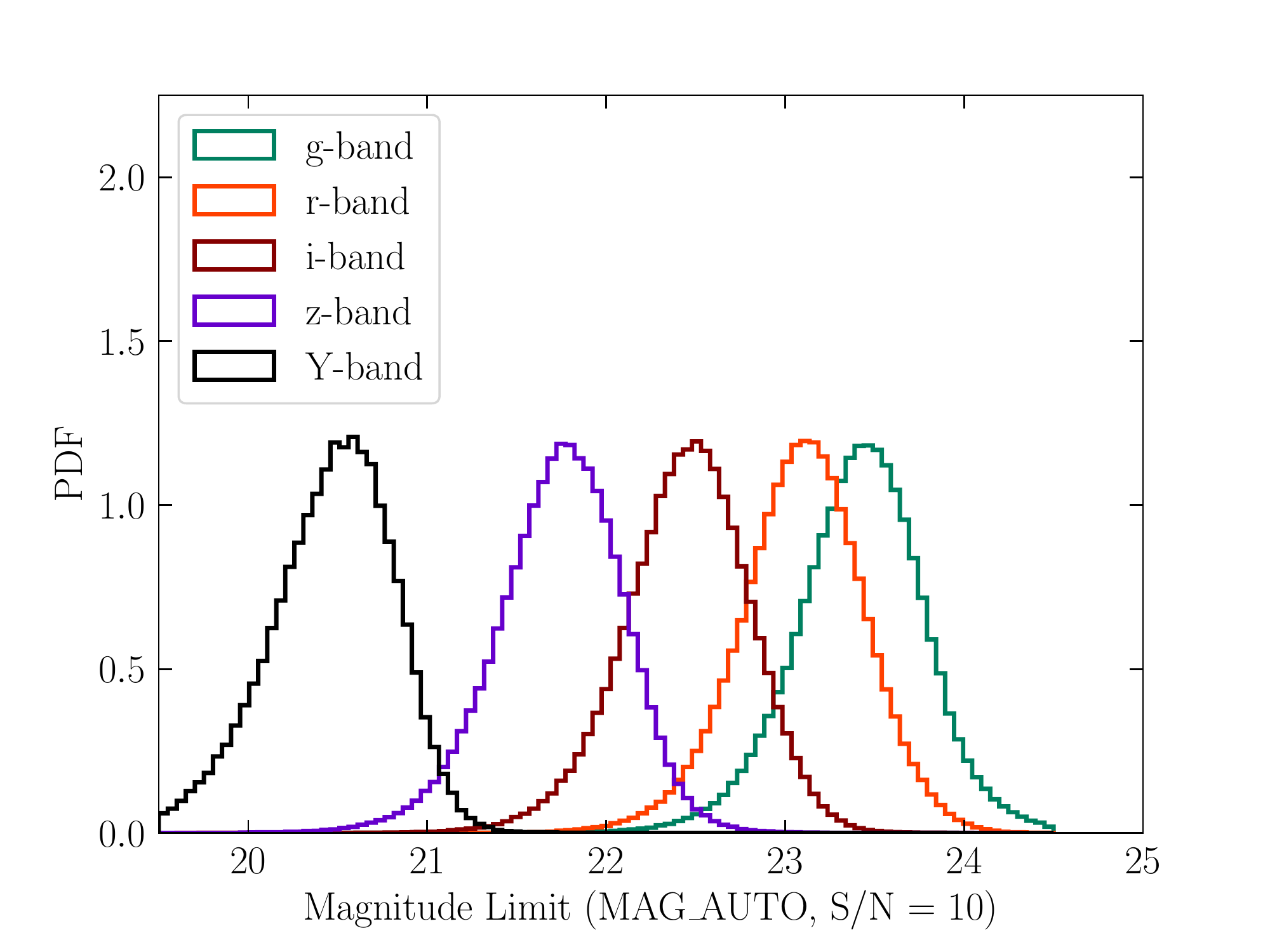}
\includegraphics[width=0.5\textwidth]{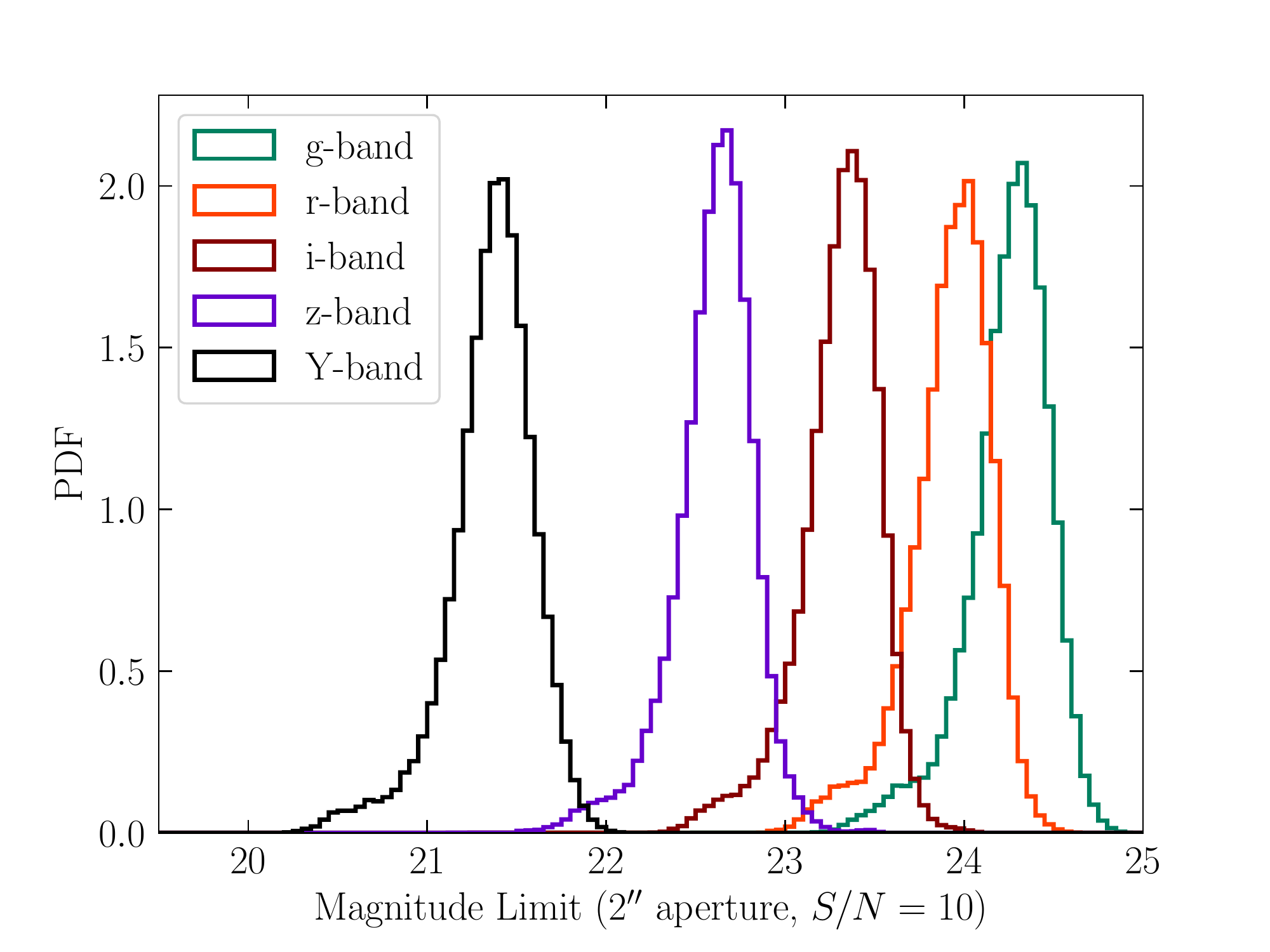}
\caption{Two estimates of the DES DR1 coadd catalog depth displayed as normalized histograms. \textit{Left}: Catalog depth estimated for \var{MAG\_AUTO} using catalog objects with ${\rm S/N} = 10$ ($\var{MAGERR\_AUTO} = 0.10857$). \textit{Right}: Catalog depth for a 2\arcsec\ aperture estimated from image properties using \mangle. \NEW{All magnitudes are given in the AB system.}}
{\label{fig:mag_magerr}\label{fig:maglim}} 
\end{figure*}

\noindent We summarize in \tabref{depth_estimates} the characteristic ${\rm S/N} = 10$ thresholds for \magauto corresponding to $\delta F/F \sim 0.1$ in each of the $grizY$ bands.
The left panel of \figref{mag_magerr} shows the \magauto distribution for DES DR1 catalog objects with $0.10837 < \var{magerr\_auto} < 0.10877$.
The spread in these distributions comes from the dependence of ${\rm S/N}$ on source properties (\eg, surface brightness) and survey non-uniformity.

\subsubsection{Depth from image properties}


It is also possible to estimate the DES DR1 imaging depth using \mangle \citep{2004MNRAS.349..115H, 2008MNRAS.387.1391S}, that generates a vectorized map of the survey coverage accounting for the focal plane geometry and imaging artifacts (i.e., bright star masks, bleed trails, satellite trails) \citep{Y1A1,DESDM}.
The \mangle processing produces a coadd weight map from a weighted sum of the single-epoch input images.
This weight was converted to a ${\rm S/N} = 10$ limiting magnitude for a $2\arcsec$ diameter aperture, corresponding approximately to the \var{MAG\_APER\_4} quantity measured by \sextractor \citep[for details, see][]{Y1A1}.
The median limiting magnitude across the DES footprint is \NEW{$g = \maglimmangg$, $r = \maglimmangr$, $i = \maglimmangi$, $z = \maglimmangz$, $Y = \maglimmangy$} (right panel of \figref{maglim}).

\subsubsection{Object detection completeness}

Another measure of effective imaging depth is the object detection completeness relative to deeper imaging data.
We evaluated the detection completeness of DES DR1 through a comparison to public CFHTLenS data \citep{2013MNRAS.433.2545E}
using an overlap region centered on $({\rm RA, DEC}) = (34\fdg5, -5\fdg4)$ consisting of 9 CFHTLenS fields in the W1 patch.\footnote{CFHTLenS object catalogs and image masks available at \url{http://www.cfhtlens.org/astronomers/data-store}}
The CFHTLens $5\sigma$ magnitude limit for a $2\arcsec$ aperture is $g' = 25.58$, $r' = 24.88$, $i' = 24.54$, $y' = 24.71$ and $z' = 23.46$.  
To ensure full coverage in both surveys, we restricted the analysis to regions with CFHTLenS image mask value $\var{mask} = 0$ and DES coverage fraction $>99\%$ in the intersection of the $griz$ bands \citep[see][]{Y1A1}.
The effective area of overlap in both surveys including masking is 6.0~deg$^2$.

Object matching is performed using a $1\arcsec$ radius and we require a robust flux measurement in the respective DES band ($15 < \magauto < 30$) that is roughly consistent with that of CFHTLenS (within $1\magn$) for an object to count as ``detected'' in DES DR1.

\begin{figure}
\includegraphics[width=0.5\textwidth]{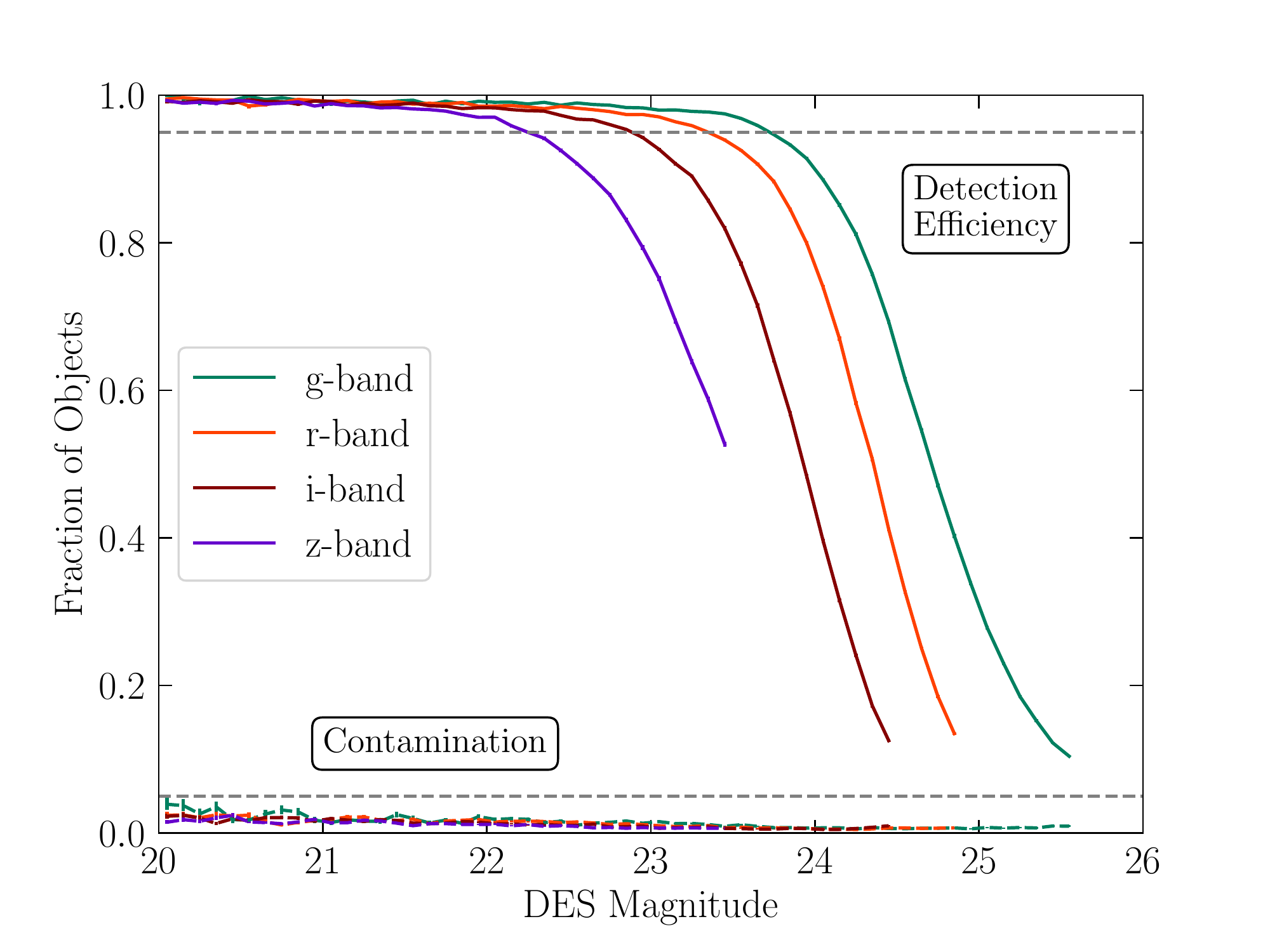}
\caption{DES DR1 detection efficiency  and contamination relative to deeper imaging from CFHTLenS. Solid color curves represent the detection efficiency, while dashed color curves show the fraction of unmatched objects appearing only in DES. For visual reference, gray dashed lines indicate 5\% and 95\% of objects. \NEW{DES magnitude is given in the AB system.}}\label{fig:detection_efficiency}
\end{figure}

The DES DR1 detection efficiency is defined as the fraction of CFHTLenS objects in a given flux interval that has a matched DES object passing the baseline quality cuts listed above, and is expressed in the DES photometric system using converted flux measurement from CFHTLenS.
The resulting detection efficiency curves are plotted in \figref{detection_efficiency}.
DES DR1 detection efficiencies are only plotted for the magnitude range brighter than the typical ${\rm S/N} = 5$ limiting magnitude of CFHTLenS.
The 95\% completeness magnitude threshold obtained from this test is $g = \magcompleteg$, $r = \magcompleter$, $i = \magcompletei$, and $z = \magcompletez$ (\tabref{summary}).
CFHTLenS does not include comparable $Y$-band coverage.

Using the same baseline quality selection criteria of $\var{flags\_[griz]} < 4$ and \var{imaflags\_iso\_[griz]} = 0, we find that in each of $griz$ bands, $\lesssim1\%$ of DES objects with \magauto greater than 20 and less than the typical ${\rm S/N} = 5$ limiting magnitude of CFHTLenS lack a matched counterpart in CFHTLenS (\figref{detection_efficiency}).
This suggests that contamination from spurious objects in DES DR1 is also $\lesssim1\%$. 
We expect that some fraction of unmatched DES objects are astrophysical transients or moving objects.
Indeed, the spatial distribution of coadd objects with only one single-epoch detection across the $grizY$ bands is concentrated along the ecliptic.


\subsection{Morphological object classification}\label{sec:sgsep}


A basic selection on object size relative to the PSF can be used to separate samples of spatially extended galaxies from point-like stars and quasars.
Accurate object classification becomes challenging for ground-based imaging surveys at faint magnitudes, and accordingly, optimal use of morphological, color, and temporal information is an active area of research \citep[e.g.,][]{2012ApJ...760...15F,2013A&A...557A..16M,2015MNRAS.453..507K, 2015A&C....10...43B, 2017MNRAS.464.4463K}.
Several object classification schemes have been applied to DES data for a variety of science cases \citep[e.g.,][]{2015MNRAS.450..666S,2015ApJ...801...73C,2015MNRAS.454.3952R,Y1A1,sgsepY1}.
The most common classification scheme makes use of the \sextractor \spreadmodel, that compares the fit of a local PSF model to a slightly extended exponential disk model \citep{2012ApJ...757...83D}.
Below we show an example using \spreadmodel for object classification in a way that is suitable for both stellar and extragalactic science.

We define a new variable \var{extended\_coadd} as the independent sum of several Boolean conditions:

\footnotesize\begin{align}\begin{split}
\var{ext}&\var{ended\_coadd} = \\
~&((\var{spread\_model\_i} + 3\, \var{spreaderr\_model\_i}) > 0.005) \\
+&((\var{spread\_model\_i} + \var{spreaderr\_model\_i}) > 0.003) \\
+&((\var{spread\_model\_i} - \var{spreaderr\_model\_i}) > 0.003).
\end{split}\end{align}
\normalsize

\begin{figure*}
\includegraphics[width=0.5\textwidth]{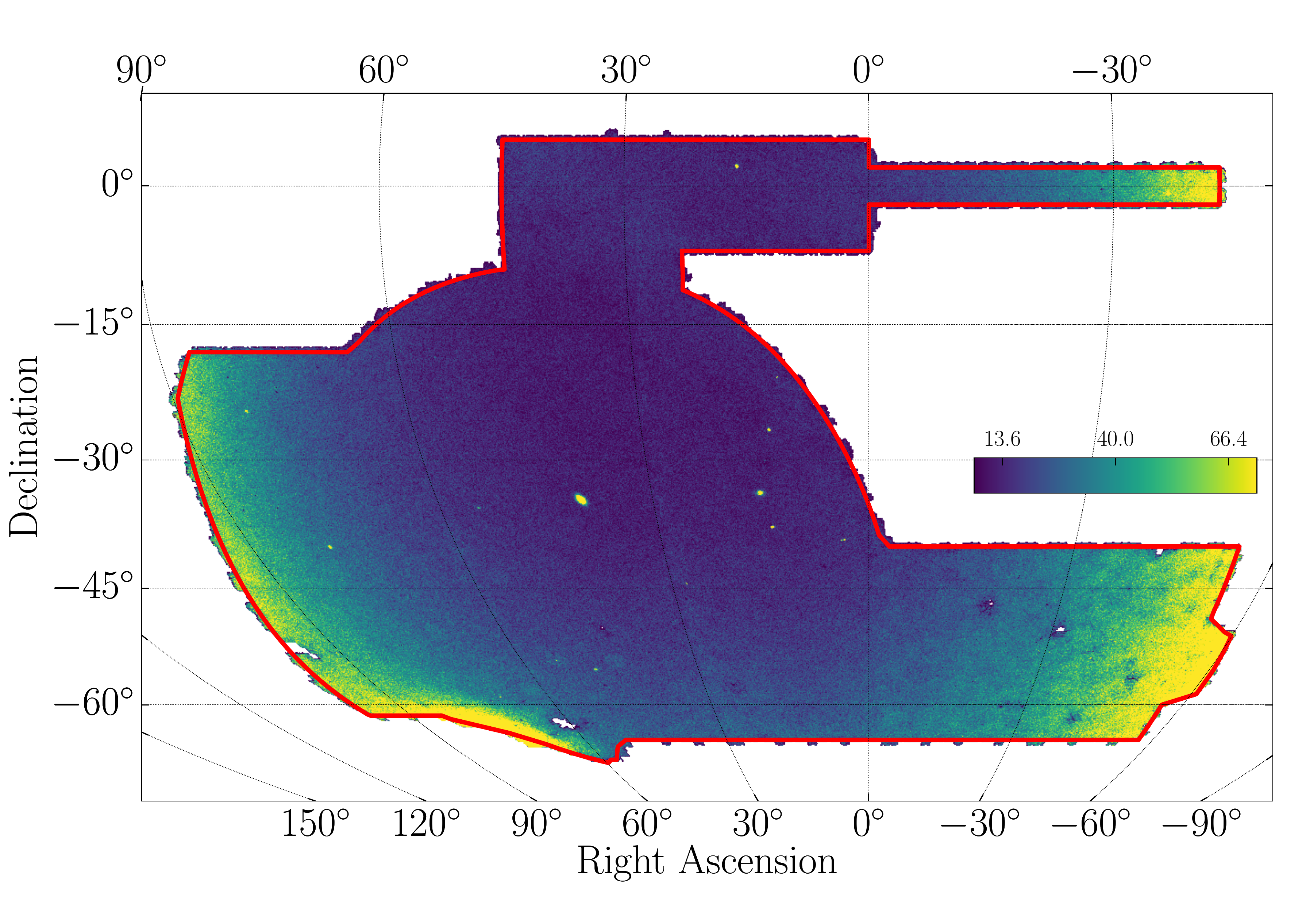}
\includegraphics[width=0.5\textwidth]{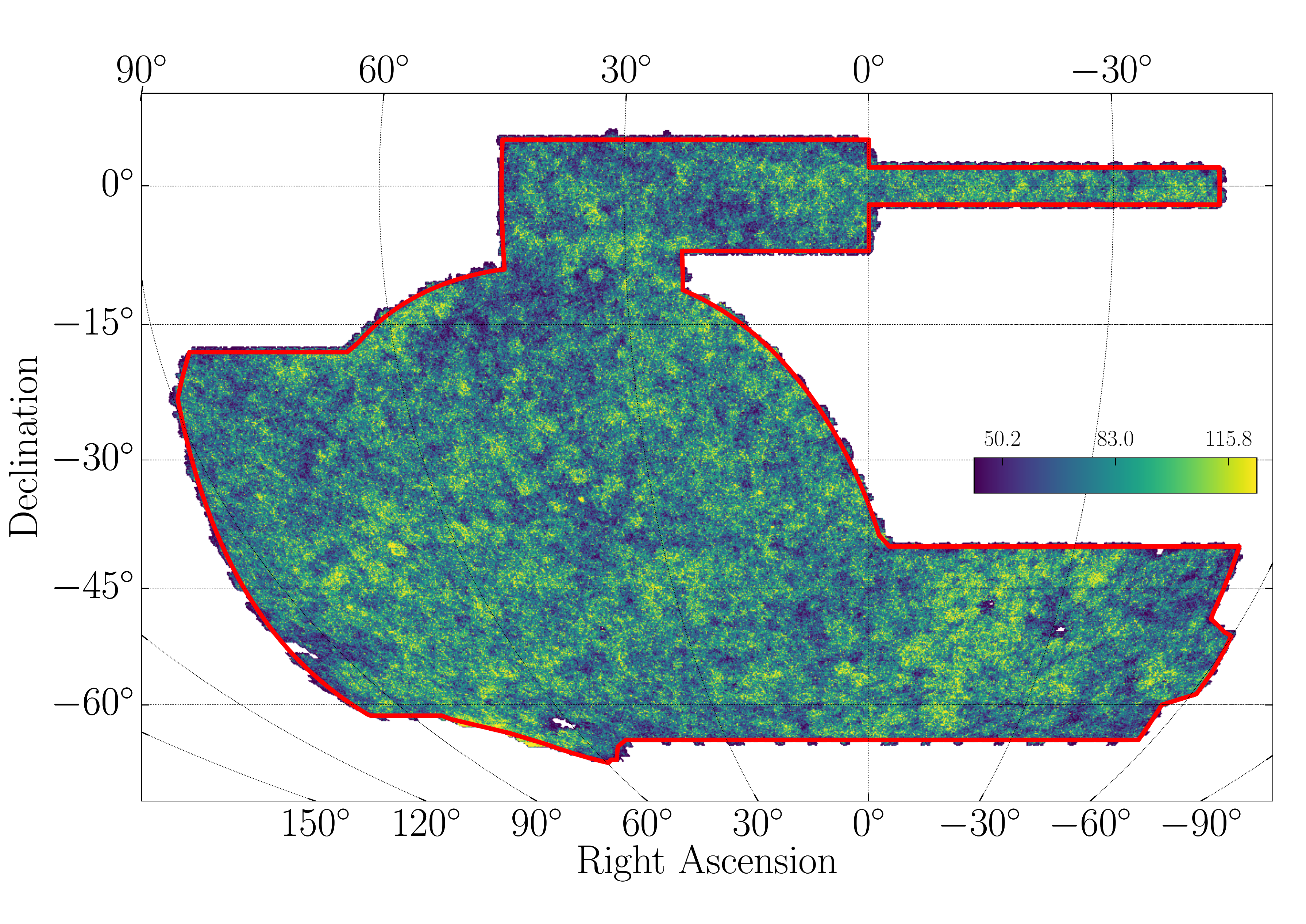}
\caption{\textit{Left}: Stellar density map at \healpix $\nside = 1024$ resolution created with the $\var{EXTENDED\_COADD} = 0$ selection described in \secref{sgsep} (see also \appref{queries}). Discrete peaks in the stellar density correspond to globular clusters and dwarf galaxies in the Milky Way halo. \textit{Right}: Analogous galaxy density map created with the $\var{EXTENDED\_COADD} = 3$ selection. \NEW{Color range units are number of objects per \healpix $\nside = 1024$ pixel.}}\label{fig:densities}
\end{figure*}

\noindent  Note that \var{extended\_coadd} is defined by a sequence of boolean conditions that, when true, add a unit to the classifier.
This classifier results in a value of 0 (high-confidence stars), 1 (likely stars), 2 (mostly galaxies), and 3 (high-confidence galaxies).
\appref{queries} demonstrates how this classifier can be used in a SQL query statement. \figref{densities} shows  density maps for stars and galaxies selected using \var{extended\_coadd} equals to $0$ and $3$ respectively.

\begin{figure}
\includegraphics[width=0.5\textwidth]{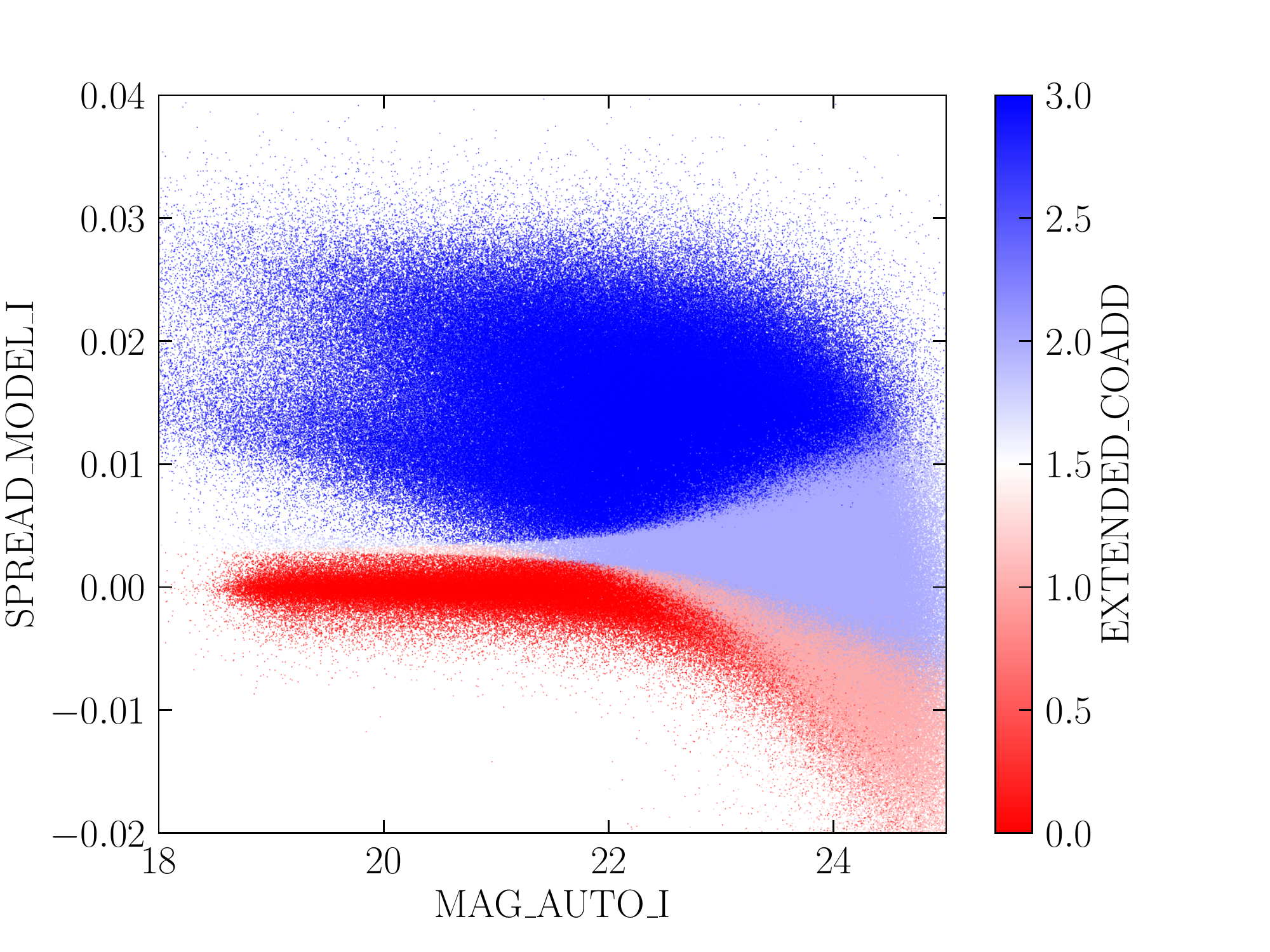}
\includegraphics[width=0.5\textwidth]{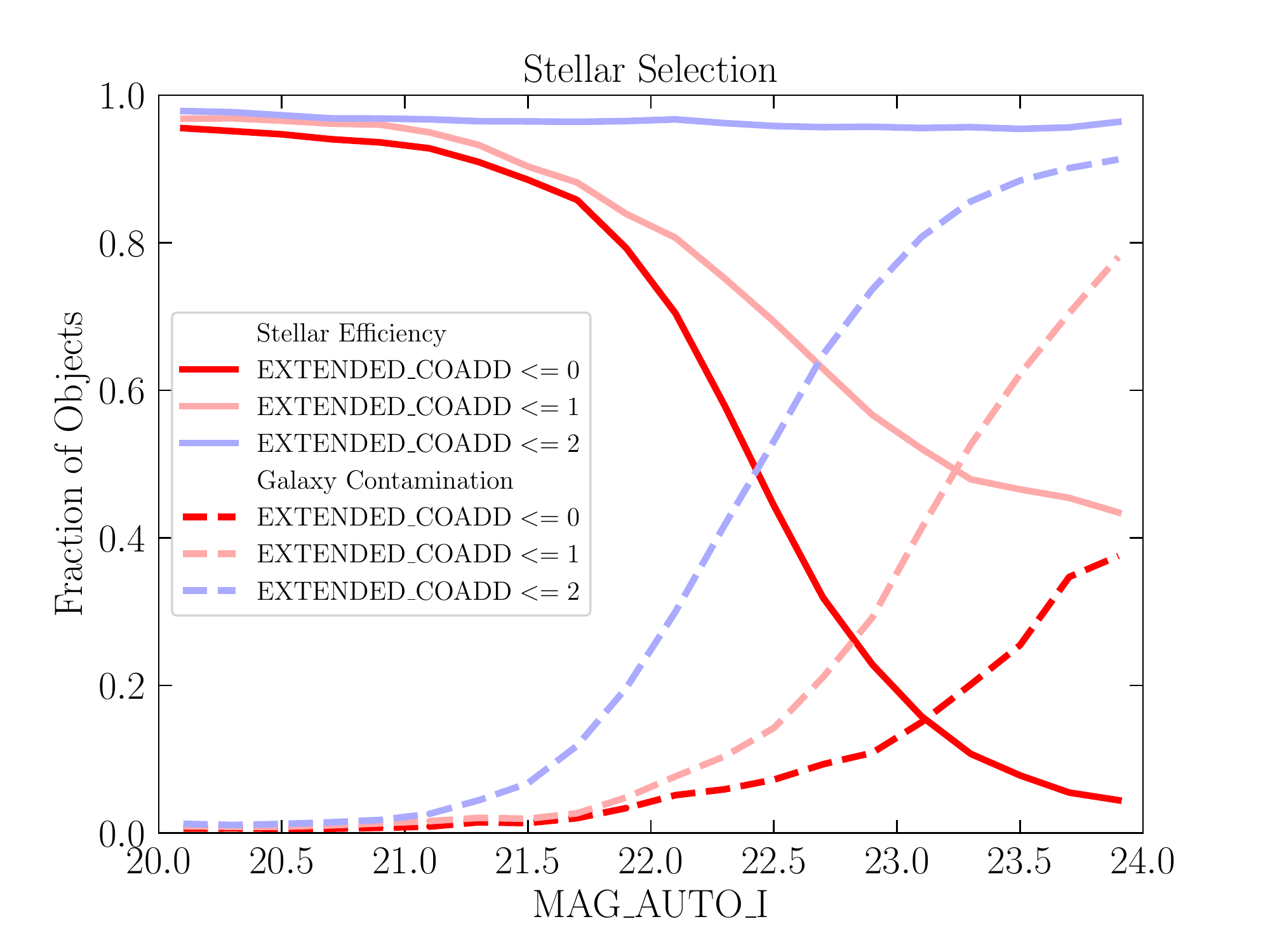}
\includegraphics[width=0.5\textwidth]{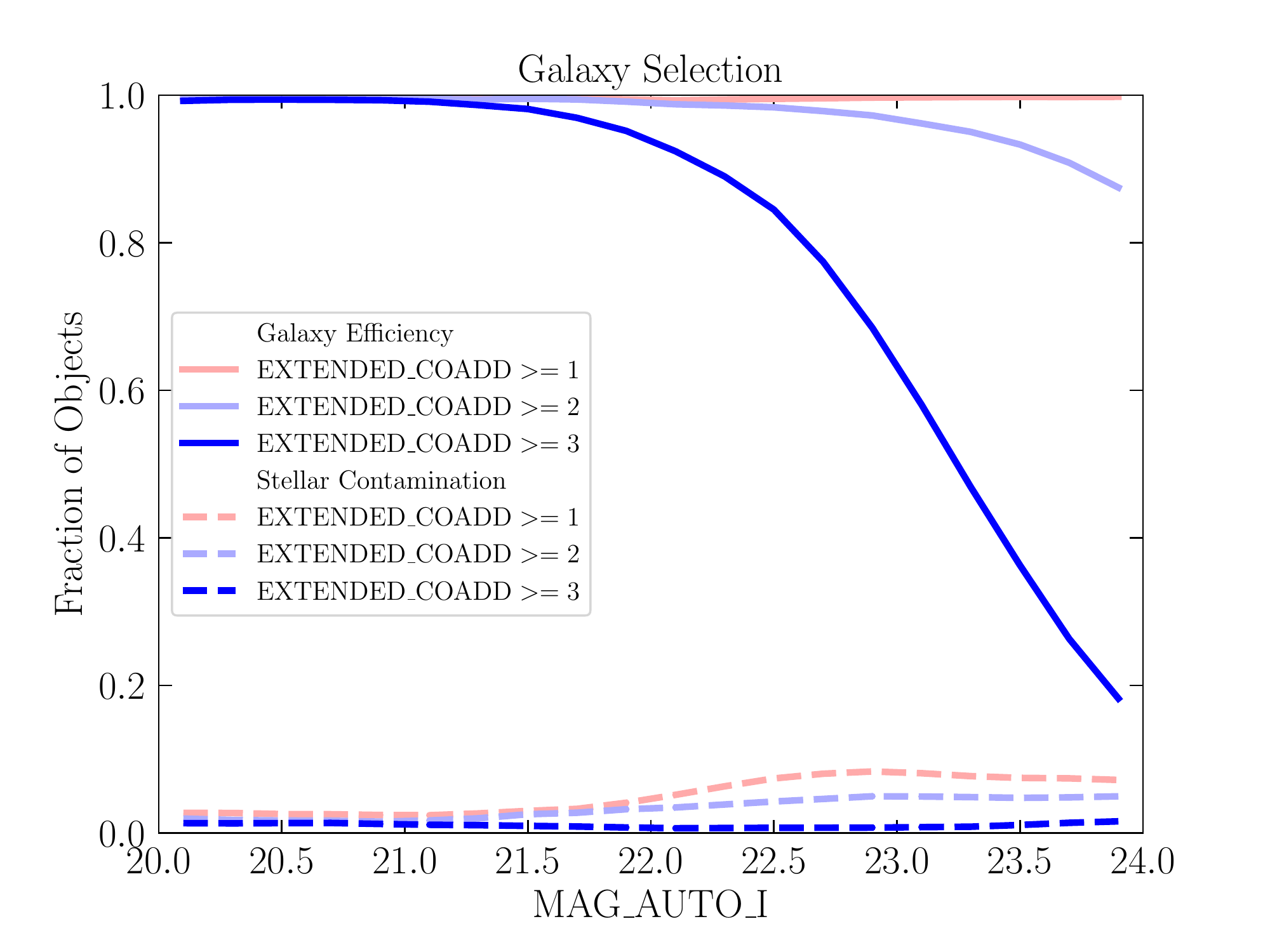}
\caption{Stars and galaxies occupy distinct regions of spreadmodel-space at bright magnitudes, but become more difficult to distinguish at faint magnitudes (\textit{top}). DES DR1 object classification accuracy versus HSC-SSP for both stellar (\textit{middle}) and galaxy (\textit{bottom}) samples. By using an interval of \var{EXTENDED\_COADD} values (\secref{sgsep}), the balance of classification efficiency and purity can be adjusted as appropriate for specific science cases. \NEW{DES magnitude is given in the AB system.}}\label{fig:sgsep}
\end{figure}

We evaluate the performance of the example classifier above using three regions in the main body of the DES footprint that overlap fields from HSC-SSP DR1 \citep{2017arXiv170208449A} with typical seeing in the $i$-band $\var{FWHM} \lesssim 0.7\arcsec$: SXDS (Ultra Deep layer), DEEP2\_3 (Deep layer), and portions of VVDS (Wide layer).
The areal overlap between these HSC-SSP data sets and DES is $\sim18$~deg$^{2}$.
The HSC-SSP data are of sufficient depth and image quality that a distinct stellar locus is clearly visible in the HSC concentration parameter $\var{imag\_psf} - \var{icmodel\_mag}$ to an $i$-band magnitude of $\roughly24.0$.
We choose empirically an interval of \var{extended\_coadd} values to select stellar or galactic samples with a balance of classification efficiency and purity appropriate for different science cases (\figref{sgsep}).
For instance, one can define galaxy ($\var{extended\_coadd} >= 2$) and stellar samples ($\var{extended\_coadd} <= 1$) having $\roughly310{\rm M}$ and $\roughly80{\rm M}$ objects, respectively, following the standard object quality selection.

\begin{figure}
\includegraphics[width=0.5\textwidth]{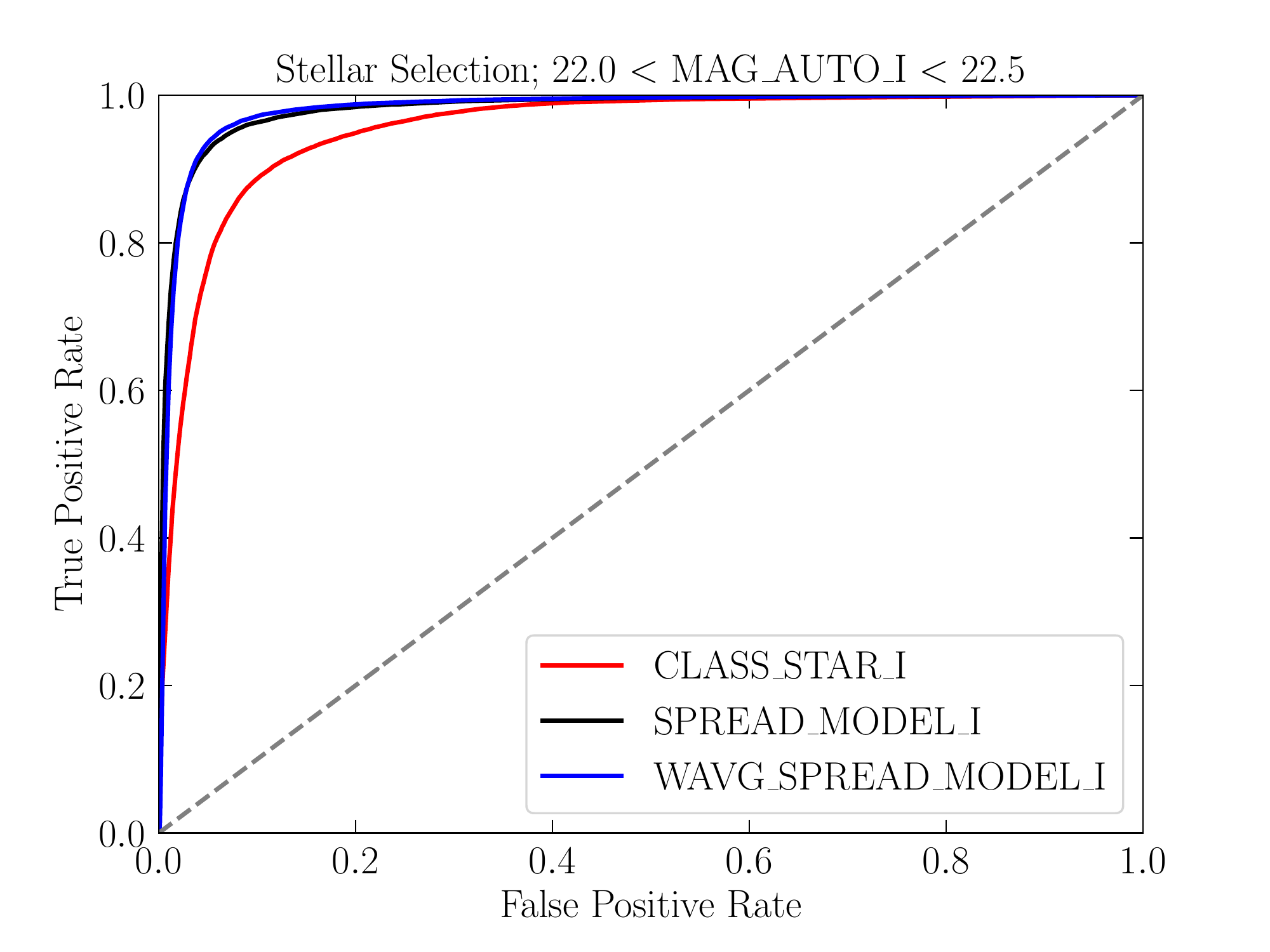}
\caption{Receiver operating characteristic (ROC) curve for a stellar selection in the magnitude range $22.0 < \magauto[i] < 22.5$ \NEW{(AB system)}. 
The \spreadmodel outperforms \classstar in classification accuracy of DES DR1 objects with respect to HSC-SSP. \NEW{In this case, the \spreadmodel and \wavgspreadmodel variants provide a very similar classification power, as denoted by the closely overlapping black and blue lines respectively. Diagonal dashed line corresponds to the expectation from a random classifier.}}\label{fig:roc}
\end{figure}

We recommend using quantities based on \sextractor \spreadmodel for morphological classification in DR1.
Although the \sextractor quantity \classstar has been commonly used in the past, we find that \spreadmodel-based classifiers consistently outperform \classstar, as exemplified by the receiver operating characteristic (ROC) curve shown in \figref{roc} and summarized in \tabref{auc}.
The ROC curves are generated by performing a simple scan of threshold values for each of \sextractor quantities, and using the HSC-SSP classifications described above as a reference.
Classifiers based on the weighted averaged \spreadmodel from single epoch detections, \wavgspreadmodel, are expected to be more robust for objects that are bright enough to be detected in single-epoch imaging (see \secref{issues}).
It is expected that classifiers using alternative bands and/or combinations of object measurements will be more appropriate for specific science cases.

\begin{\tabletype}{l c c c}
\tablewidth{0pt}
\tabletypesize{\tablesize}
\tablecaption{Stellar classification accuracy quantified as the area under the ROC curve (AUC) in three flux intervals, using HSC-SSP as a reference (see \figref{roc}). In this case, the AUC statistic is the probability that the simple classifier will correctly rank a randomly chosen star higher than a randomly chosen galaxy. \NEW{DES magnitude is given in the AB system.} \label{tab:auc}}
\tablehead{
Quantity & \wavgspreadmodel & \spreadmodel & \classstar \\
}
\startdata
$21.5 < \magauto[i] < 22.0$ & 0.994 & 0.991 & 0.979 \\
$22.0 < \magauto[i] < 22.5$ & 0.981 & 0.981 & 0.954 \\
$22.5 < \magauto[i] < 23.0$ & 0.917 & 0.948 & 0.881 \\
\enddata
\end{\tabletype}

\subsection{Known issues}\label{sec:issues}

\begin{figure}
\includegraphics[width=0.5\textwidth]{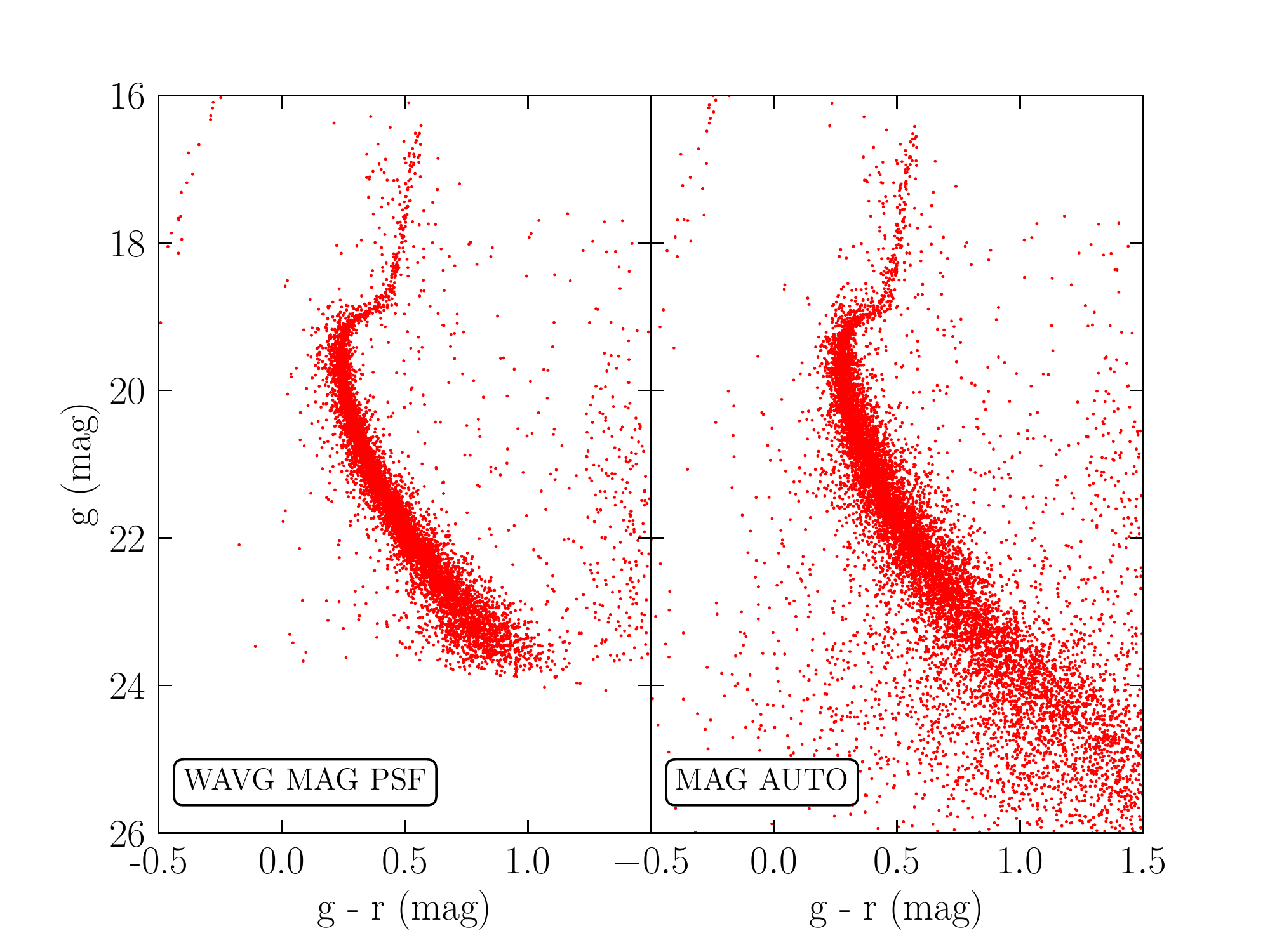}
\caption{Color-magnitude diagram for a stellar sample selected within a $15\arcmin \times 15\arcmin$ box centered on the M2 globular cluster (see \appref{queries}). The weighted-average PSF photometry from \var{wavg\_mag\_psf} (\textit{left}) yields a tighter locus but does not extend as deep as the \magauto photometry (\textit{right}). \NEW{All magnitudes are given in the AB system.}}\label{fig:globular_cluster}
\end{figure}

The PSF model has limited flexibility to accommodate discontinuities in the effective PSF that can occur in coadded images at boundaries in coverage between individual exposures.
In these regions, the local PSF model can fail to accurately fit the point-like sources, and accordingly, both the coadd morphological classifications (\ie, \spreadmodel, \var{class\_star}) and coadd PSF photometry are suspect.
The photometry of extended sources is impacted to a lesser degree.
Using the $i$-band as an example, we identified such coadd PSF failures by searching for regions with anomalous coadd \spreadmodel[i] distributions and estimate that $\lesssim0.4\%$ of the footprint is substantially affected.
\KB{Estimated from y3a2\_badregions\_mask\_v1.0.fits.gz available at https://cdcvs.fnal.gov/redmine/projects/des-y3/wiki/Y3A2\_Gold\_bad\_regions.}
Generally, more robust treatment (especially for point-like sources) is possible with weighted-average quantities, at the cost of some loss of object detection completeness for the faintest sources (\figref{globular_cluster}).
Due to the known issues in coadd PSF photometry, DES DR1 does not include the coadd \var{MAG\_PSF} quantities. 
For studies of point-like sources, we recommend the use of \var{WAVG\_MAG\_PSF} (bright sources) or \var{MAG\_AUTO} (faint sources).
\ADW{We should have a list of quantities that are most affected: \var{CLASS\_STAR}, \var{SPREAD\_MODEL}, model magnitudes?}


\begin{figure}
\includegraphics[width=0.48\textwidth]{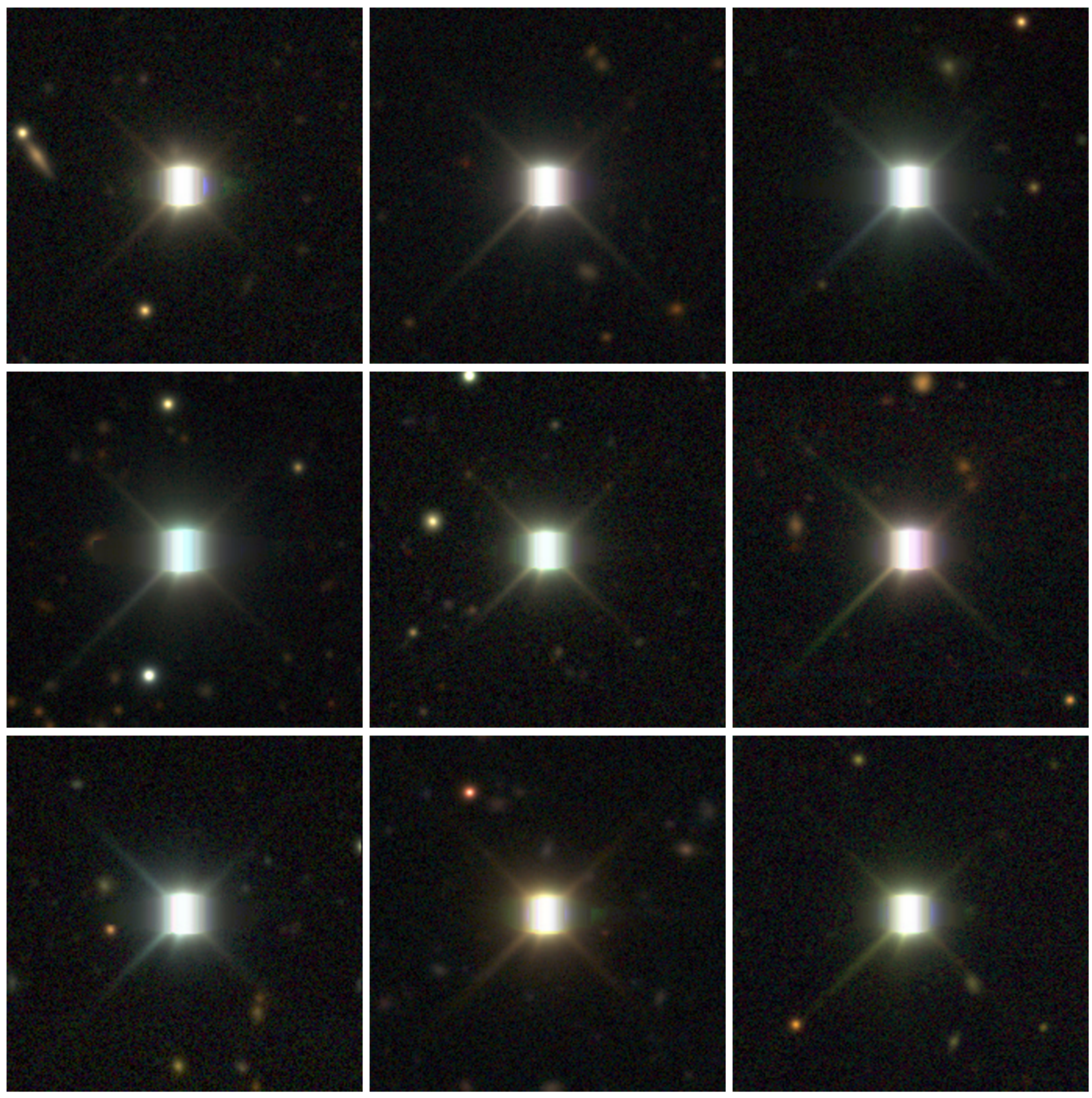}
\caption{Composite $gri$ $1\arcmin \times 1\arcmin$ cutouts of saturated stars with \imaflagsiso = 1.}\label{fig:saturated}
\end{figure}

\begin{figure}
\includegraphics[width=0.48\textwidth]{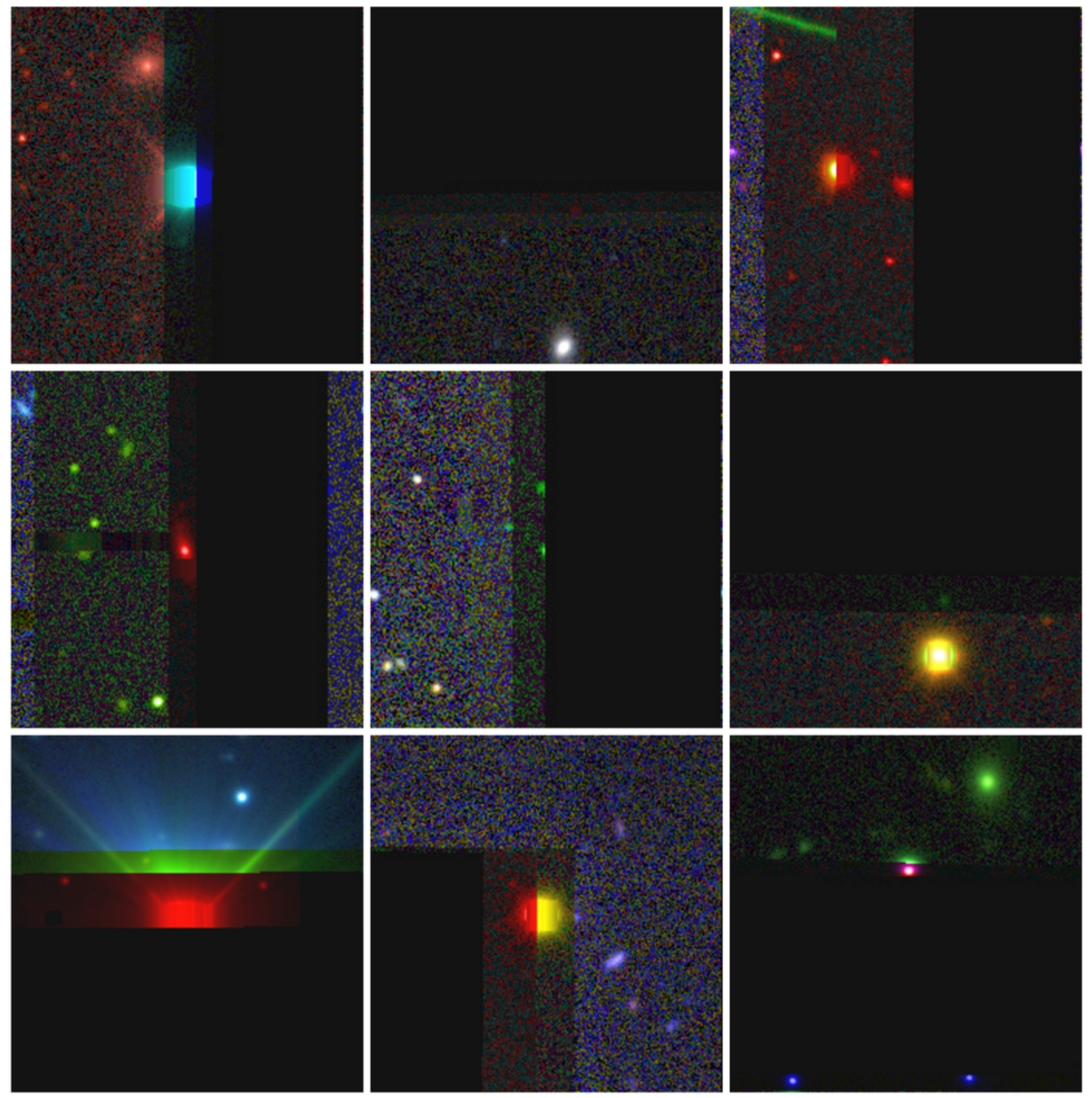}
\caption{Composite $gri$ $1\arcmin \times 1\arcmin$ cutouts of objects with \imaflagsiso = 1 and \niter = 0. The majority of such objects, but not all,
have missing imaging data.}\label{fig:niter}
\end{figure}

Among all the objects detected and cataloged, $\roughly 2.3\%$ have been flagged by \var{IMAFLAGS\_ISO} in at least one of the $grizY$ bands, and $\roughly 0.1\%$ have artifacts in all 5 bands. As described in \secref{flags} it is recommended to use $\imaflagsiso=0$ as a first filter of clean objects. 
Most of the objects with $\imaflagsiso=1$ are saturated stars (\figref{saturated}). 
A smaller fraction of flagged objects are missing imaging data in one or more of the bands, as shown in \figref{niter}, and usually have $\niter = 0$, that is set for objects that did not converge during the photometry measurement.

\begin{figure}
\includegraphics[width=0.48\textwidth]{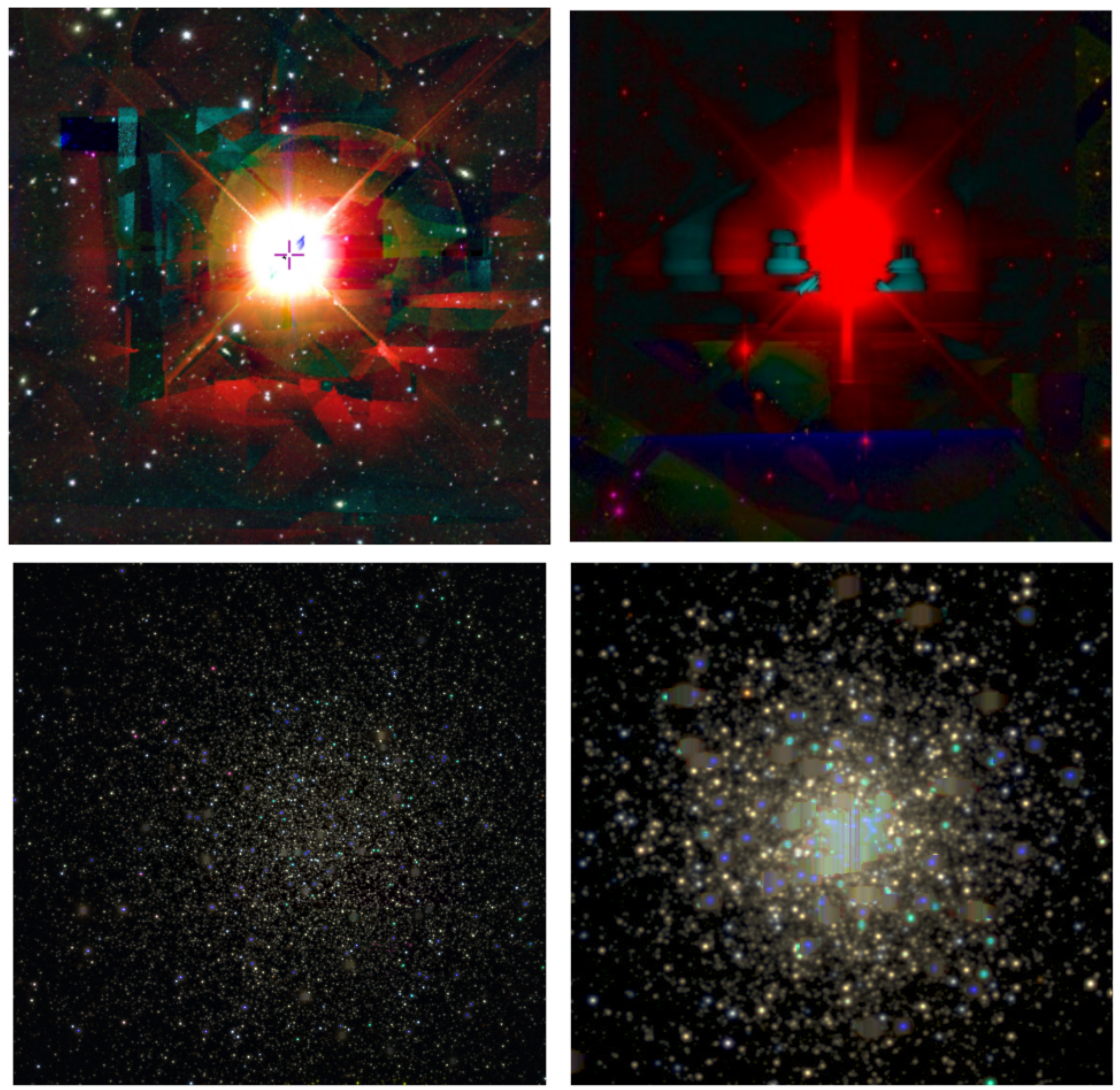}
\caption{Examples of very bright stars, $R$ Dor ($30\arcmin \times 30\arcmin$ cutout; \textit{top left}) and $\alpha$ Phe ($14\arcmin \times 14\arcmin$ cutout; \textit{top right}) and globular clusters, NGC 288 ($8\arcmin \times 8\arcmin$ cutout; \textit{bottom left}) and NGC 1904 ($2\farcm5 \times 2\farcm5$ cutout; \textit{bottom right}), found within the DES DR1 footprint.}\label{fig:bright}
\end{figure}

Scattered light from very bright stars can impact the photometry of nearby objects.
\tabref{bright_stars} lists the coordinates and magnitudes of bright stars in the DES footprint. 
These stars generally cause ragged holes in the imaging coverage of DES DR1 due to image-level blacklisting of scattered light artifacts (\figref{bright}). 
Scattered light from these stars extends beyond image- and catalog-level masking, and can be observed at $> 1\degree$ in the number counts of objects with extreme colors, \ie $(g-r) > 4$ or $(i - z) > 4$ \citep{Y1A1}.  
Additional care must be taken for analyses close to bright stars.

The DESDM processing pipeline is designed for extra-galactic science at high-Galactic latitudes.
As such, it is not optimized to detect or measure sources in extremely crowded regions.
Failures in source deblending are most noticeable in the cores of Galactic globular clusters, where source confusion and saturation reduce the catalog completeness appreciably.
\tabref{bright_gc} lists 5 classical globular clusters within the DES DR1 footprint with integrated $V$-band magnitude brighter than 10 \citep[][2010 edition]{1996AJ....112.1487H}.
The bottom panels of \figref{bright} show two of these clusters, NGC~288 and NGC~1904, where saturation and source confusion can greatly impact source detection and photometric measurements. 
In addition, nearby bright galaxies (such as NGC~253, NGC~247, NGC~55, and IC~1613) are sometimes deblended into numerous individual objects, and some nearby edge-on spiral galaxies are partially masked.


\begin{deluxetable}{l c c c }
\tablehead{\colhead{Star} & \colhead{RA} & \colhead{DEC} & \colhead{Apparent Magnitude} \\ 
  & (deg) & (deg) & (V mag) }
\tablecaption{List of very bright stars within the DES footprint. \label{tab:bright_stars}}

\startdata
$\alpha$ Col (Phact)    & 84.9121  & -34.0741 & 2.65\\
$\alpha$ Phe (Ankaa)    & 6.5708   & -42.3061 & 2.38\\
$\alpha$ Eri (Achernar) & 24.4288  & -57.2367 & 0.46 \\
$\alpha$ Hya (Alphard)  & 29.6925  & -61.5697 & 2.00\\
$\gamma$ Eri (Zaurak)   & 59.5075  & -13.5086 & 2.91\\
$R$ Dor                 & 69.1900  & -62.0775 & 5.40\\
$\alpha$ Car (Canopus)  & 95.9879  & -52.6958 & -0.74\\
$\alpha$ Pav (Peacock)  & 306.4121 & -56.7350 & 1.94\\
$\alpha$ Gru (Alnair)   & 332.0583 & -46.9611 & 1.74\\
$\beta$  Gru (Tiaki)    & 340.6671 & -46.8847 & 2.15\\
\enddata
\end{deluxetable}

\begin{deluxetable}{l c c c }
\tablehead{\colhead{Globular Cluster} & \colhead{RA} & \colhead{DEC} & \colhead{Apparent Magnitude} \\
 & (deg) & (deg) & (mag) }
\tablecaption{Brightest globular clusters within DES DR1. \label{tab:bright_gc}}
\startdata
NGC 288	 &      13.2	& -26.58 & 8.1	\\
NGC 1261 &	48.075   &-55.13 & 8.4	\\
NGC 1851 &	78.525	 & -40.05 & 7.3	\\
NGC 1904 &	81.045	& -24.523 & 8.0	\\
NGC 7089 &	323.375	& -0.8167 & 6.5	\\
\enddata
\end{deluxetable}

The tabulated ``DES DR1 Standard Bandpasses'' (\figref{filters}, \secref{bandpass}) included in this release were not the ones used to calculate the zeropoints applied to the calibrated images and coadd source catalogs presented in DR1, but instead correspond to an earlier version of the system throughput referred to as the ``Y3A1 Standard Bandpasses''. The latter system, that was the one available at the time of data processing, differs from the DES DR1 Standard Bandpasses in the treatment of the out-of-band system response.
The DES DR1 Standard Bandpasses correct for a small inaccurate representation of $r$-band throughput in the Y3A1 Standard Bandpasses, which due to an incorrect input calibration run, 
considered an unrealistically large light leakage relative to the in-band response ($\sim10^{-2}$) at wavelengths $\roughly8000\angstrom$ and $\roughly9200\angstrom$.
For coadd objects, the impact of setting the out-of-band response to zero in the DES DR1 Standard Bandpasses would lead to a photometry difference of $\lesssim2$~\mmag (RMS) in the $grizY$ bands, which is below the level of statistical uncertainty in the coadd zeropoints (\secref{photometry}).

\begin{figure}
\includegraphics[width=0.48\textwidth]{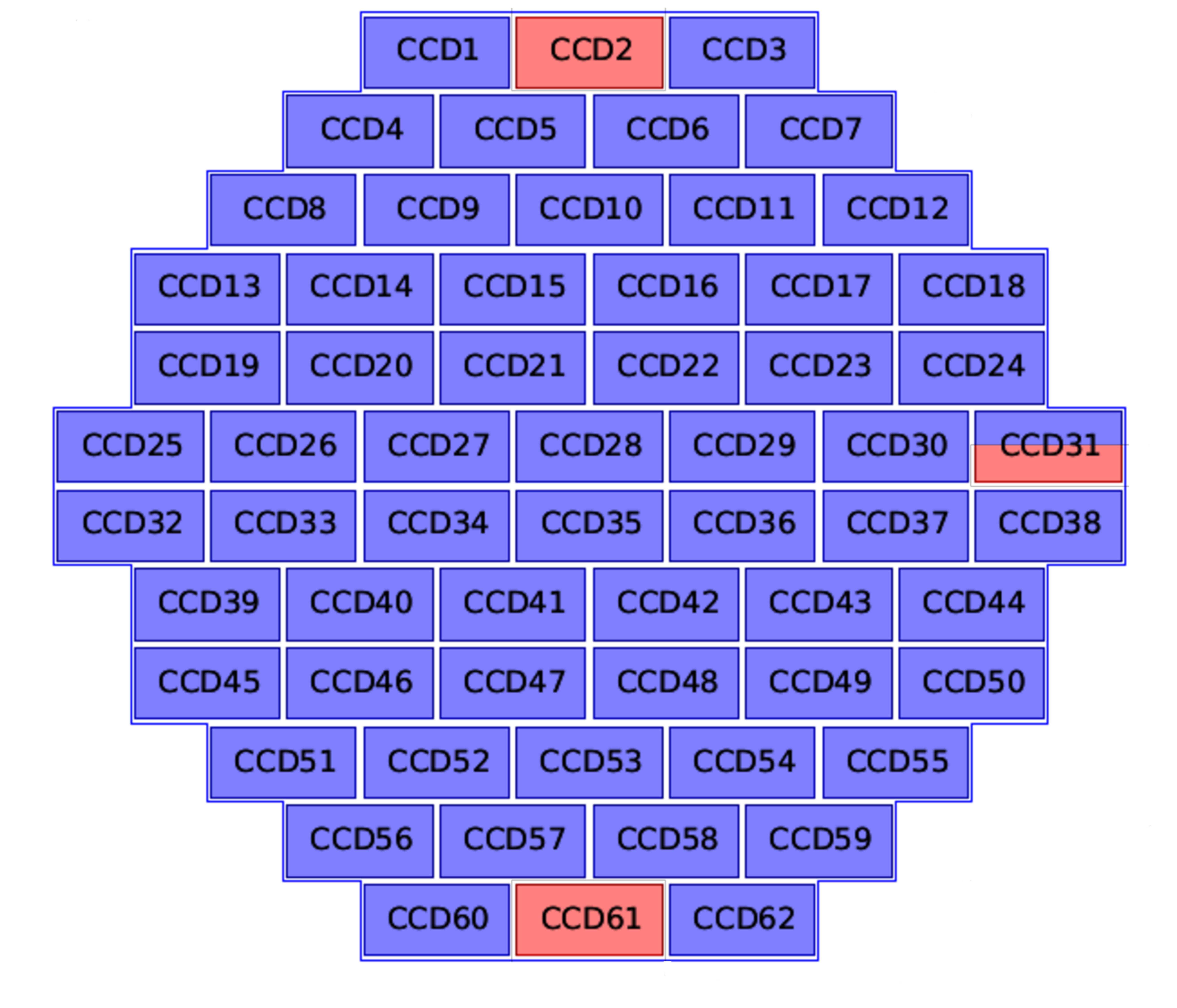}
\caption{\NEW{DECam focal plane CCD layout oriented with north at the the top and east on the right. The rectangles represent the 62 science CCDs, each of which is divided into two halves oriented along the long direction and read by two amplifiers. CCDs 2 and 61 were inactive for most of the DES data included in DR1. Amplifier A of CCD 31 has time-variable gain and has not been processed for this release. These areas are marked in red.}}\label{fig:chips}
\end{figure}

\NEW{59.5 of the 62 science CCDs in the DECam focal plane have been fully operational during the DR1 data collection period.
CCD 61, which failed in 2012, was not processed for the DR1 dataset. 
Similarly, CCD 2 was not processed between November 2013 and December 2016 (overlapping with observations included in this release), during which time it was not functional. 
Amplifier A of CCD 31 has an unpredictable, time-variable gain and is not included in this release. 
Amplifier B of CCD 31 functions normally and has been included. 
The rest of the science CCDs are performing within specifications and are usable for science. See \figref{chips} for a layout of the DECam focal plane mosaic with the positions of affected CCDs marked.}

\section{Release products}
\label{sec:products}


Here we detail the individual products included in DES DR1. 
The primary components of DES DR1 are derived from the union of 10,338 coadd tiles covering the DES footprint.
The tile distribution for a portion of the SDSS Stripe 82 region is shown in the inset of \figref{footprint}. To view all tiles and a \healpix map (\nside = 32) of 1753 pixels covering the entire DR1, visit \url{https://des.ncsa.illinois.edu/easyweb/footprint}. 

\subsection{Images}
DES DR1 images can be grouped into two categories:

\begin{itemize}
    \item[a)]  Calibrated Single Exposures: 38,850 exposures with photometric calibration, corresponding to
           4,124,753  individual CCD images. The number of exposures per band are $N_g=7626$, $N_r = 7470$, $N_i = 7470$, $N_z = 7753$ and $N_Y = 8531$. Each raw exposure is $\roughly 0.5$ \GB in size (compressed). 
           These images can be accessed through NOAO Data Lab.\footnote{\url{http://datalab.noao.edu/}}
    \item[b)]  Coadd Images: As a result of the Multi-Epoch Pipeline described in \secref{preprocessing} a total of 10,338 tiles of 10k $\times$ 10k pixels spanning $0.7306 \deg$ on a side at a resolution of 0.263 \asec/pixel (see inset of \figref{footprint}) were produced in each of the five bands. These images along with the $r + i + z$ combined detection image used are publicly available with a total of 62,028 images constituting $\roughly 11$ \TB of data.
\end{itemize}
\subsection{Catalogs} 

The coadd source extraction process detected and cataloged 399,263,026 distinct objects. 
Morphological object information includes object centroids, shape parameters, \healpix indices, and processing flags.
Several different photometric measurements and associated uncertainties are provided, including \var{AUTO}, \var{PETRO}, \var{WAVG\_PSF}, and assorted aperture magnitudes (\tabref{apertures}).
These measurements are distributed in three database tables: \drmain, \drmagnitude, and \drflux served from an Oracle database at NCSA. 
The \drmain table contains all object information that is not a photometric measurement or uncertainty, augmented by \magauto and \code{WAVG\_MAG\_PSF} (and associated uncertainties), and interstellar extinction corrected versions. 
The other two tables contain auxiliary magnitude and flux measurements (with associated uncertainties) in addition to extra information that is present in all three tables, such as coordinates, flags, and \healpix indices. 
We note that these tables contain two sets of coordinates for the objects, namely, (\var{RA},\var{DEC}) and (\var{ALPHAWIN\_J2000},\var{DELTAWIN\_J2000}) that are computed in the same manner; the difference between them is that \var{RA} and \var{DEC} are truncated to six decimals in order to provide a search indexing and table partitioning on these columns, while \var{ALPHAWIN\_J2000} and \var{DELTAWIN\_J2000} are double precision quantities to be use when precise measurements are needed. 
All spatial-based queries should use \var{RA} and \var{DEC} in their condition statements.  
Additionally, the table \drtileinfo contains information about the processed tiles, such as sky location, geometry, number of objects, and file paths to access the associated images and object catalogs. 
For a complete description of these tables, we refer the reader to \appref{appendix_tables}.

\begin{deluxetable}{l|c|c }
\tablehead{\colhead{Column Name} & \colhead{Diameter} & \colhead{Diameter} \\ [-0.2em]
\colhead{} & \colhead{(pixels)} & \colhead{(\asec)}
}
\tablecaption{Diameters for the set of aperture magnitudes. \label{tab:apertures}}
\startdata
\magaper[1] & 1.85 & 0.49\\
\magaper[2] & 3.70 & 0.97\\
\magaper[3] & 5.55 & 1.46\\
\magaper[4] & 7.41 & 1.95\\
\magaper[5] & 11.11 & 2.92\\
\magaper[6] & 14.81 & 3.90\\
\magaper[7] & 18.52  & 4.87\\
\magaper[8] & 22.22 & 5.84\\
\magaper[9] & 25.93 & 6.82\\
\magaper[10] & 29.63 & 7.79\\
\magaper[11] & 44.44 & 11.69\\
\magaper[12] & 66.67 & 17.53\\
\enddata
\end{deluxetable}

\subsection{Files} 

In order to provide a convenient way to download all the catalog data at once, and noting that the tile is the basic processing unit for the survey, we have created FITS file versions of the catalog tables grouped by coadd tile.
This amounts to 31,014 total files (for the \drmain, \drmagnitude, and \drflux tables) with almost 2.5 \TB of catalog data. 
Both the catalog and corresponding image file paths can be obtained from the \drtileinfo table (see \appref{queries} for an example) and can be accessed through the interfaces described in \secref{access}. \ADW{I don't know what ``the interfaces'' means. I think we need to be more explicit.}

\subsection{DES DR1 Standard Bandpasses}\label{sec:bandpass} 

DES DR1 includes an updated characterization of the Blanco/DECam total system response (instrument and atmosphere) for the $grizY$ bands (\figref{filters}).\footnote{\url{http://www.ctio.noao.edu/noao/content/DECam-filter-information}} 
The DES DR1 Standard Bandpasses are defined as the average CCD response across the focal plane as measured with the DECal system \citep{2013arXiv1302.5720M}, together with a standard atmospheric transmission computed with the MODTRAN IV code \citep{1999SPIE.3756..348B} using parameters typical of the environmental conditions encountered during DES observations \citep[for details, see][]{2017arXiv170601542B}.
For example, the airmass adopted for the standard atmosphere, 1.2, is well matched to the median airmass of exposures entering the coadd, 1.22.
The system response is defined in steps of 5 \angstrom from 3800 \angstrom to 11000 \angstrom.
Out-of-band light leakage has been directly measured with DECal to be $\lesssim10^{-3}$ relative to the in-band response, and vendor measurements of witness samples suggest that the out-of-band leakage is typically at the $10^{-5}$ to $10^{-4}$ level.
While detailed characterization of the out-of-band response is ongoing, the throughput of the DES DR1 Standard Bandpasses are defined as zero for out-of-band wavelengths (caveats are mentioned in \secref{issues}).



\subsection{Software} 
All software used in the DESDM pipelines described in \secref{preprocessing} and in \citet{DESDM} can be accessed from the release page itself\footnote{\url{https://des.ncsa.illinois.edu/releases/dr1/dr1-docs/processing}} or from the DES Github Organization.\footnote{\url{https://github.com/DarkEnergySurvey}} 
Together with access to the software used to generate these products and the configuration described in \citet{DESDM}, we provide the main ingredients to reprocess the data in a manner similar to that done by DES. \ADW{Do we want to say something like ``but we do not promise to support or maintain these products''?}

\section{Data Access}\label{sec:access}
Access to the DES DR1 data is provided through a collaborative partnership between NCSA,\footnote{\url{http://www.ncsa.illinois.edu/}} LIneA,\footnote{\url{http://www.linea.gov.br/}} and NOAO.\footnote{\url{https://www.noao.edu/}} From these
institutions, a rich and complementary set of tools and interfaces were developed to access and interact with DES data in
different ways that cover a broad set of use cases that enable scientific discovery.
In this section we describe the main tools provided to access DES DR1.\footnote{\url{https://des.ncsa.illinois.edu/releases/dr1/dr1-access}}

\subsection{NCSA DESaccess}
NCSA provides the primary set of basic web applications to access DES DR1 data. Developed at NCSA, \code{DESaccess} (\url{https://des.ncsa.illinois.edu/easyweb}) provides the user with a interface to submit asynchronized jobs  to perform SQL queries against an Oracle DB that contains the DES DR1 catalogs, and to generate cutouts from a given list of positions from the coadd-images. It also contains information regarding the catalogs, an interactive footprint finder to locate positions and tiles, and means to access images and catalogs in a tile-based format. The main services can be summarized as follows:

\begin{itemize}
\item SQL web client: We provide a SQL web client that allows the user to submit asynchronous query jobs against the Oracle 12 database that contains the DES DR1 tables. The submitted jobs enter into a queue and results can be retrieved at later times in either \code{csv}, \code{FITS} \citep{1981A&AS...44..363W} or \code{HDF5} \citep{hdf5} file format, supporting compression in some of the cases. Big jobs are divided into chunks of 1.5GB to facilitate file transfer. This interface also provides means to check syntax and to evaluate synchronous jobs for a quick view. Results files are served from a HTTP server that allows file transfer to remote machines. The query used and the results persist for a limited period of time under a ``My-Jobs'' tab. The query interface is powered using \code{easyaccess} \citep{easyaccess},\footnote{\url{https://github.com/mgckind/easyaccess}} an enhanced SQL command line interpreter designed for astronomical surveys and developed for DES. 

\item Cutout server: \NEW{The cutout server allows the user to generate cutouts up to 12\arcmin\ on a side centered on a given set of positions.}
The returned files include cutouts in all bands, a cutout in the detection image, a color TIFF image created with STIFF \citep{2012ASPC..461..263B} by combining the $gri$ bands, and a PNG image. \NEW{The typical sizes for these files are 85 MB for the FITS images, 9 MB for the color TIF, and 16 MB for the PNG version, for the case of a $12\arcmin\ \times 12\arcmin\ $ cutout.}
These cutout jobs also go into a queue and results files can be retrieved later and served remotely. The header of the cutouts is a copy from the original header from the images with extra keys indicating the center of the cutout \var{RA\_CUTOUT} and \var{DEC\_CUTOUT}. 
Results are preserved for a limited period and can also be retrieved under a ``My-Jobs'' tab within the web service. 
No stitching is performed for objects near the edge of the tile. This feature might be added in the future.

\item Footprint: A light-weight interactive tool that displays the DES footprint and its tiles. This tool can be used to search for a position in the sky and return the corresponding tile information including name, corner coordinates, and list of files for download. It also allows searches by \var{TILENAME} (the tile name identifier) when the name is available. This service also provides access to all the processed tiles, including associated images and catalogs, directly from the file server.

\end{itemize}

\code{DESaccess} code\footnote{\url{https://github.com/mgckind/desaccess}} as well as the release page\footnote{\url{https://github.com/mgckind/des\_ncsa}} are open sourced. 
The frontend was developed using web components, HTML, JS and Polymer\footnote{\url{https://www.polymer-project.org/}} that allows the re-use of already existing elements. 
The backend was mainly developed using the Python Tornado web-framework.\footnote{\url{http://www.tornadoweb.org/en/stable/}} 
The submission jobs are handled by Celery,\footnote{\url{http://www.celeryproject.org/}} which is a distributed queue manager written in Python. 
All of these applications are containerized using Docker\footnote{\url{https://www.docker.com/}} and all the node resourcing and scheduling, as well as all the deployment, is managed by Kubernetes,\footnote{\url{https://kubernetes.io/}} that allows to run, manage, and scale containerized applications in a robust and efficient manner.

\subsection{LIneA Science Server}
The images and catalogs generated by DESDM for DR1  can also be accessed by an interface developed by the Laboratorio Interinstitucional de e-Astronomia (LIneA) that supports the participation of Brazilian scientists in DES. It has been designed to offer ways to examine both images and catalogs, compare these with the results of other surveys, and examine the results of queries to the database. The interface consists of the following  three services:

\begin{itemize}

\item  Sky/Image viewer: The Sky/Image  viewer combination  integrates  third-party tools to allow the user to visualize the entire sky map produced  by DES (Aladin Light\footnote{\url{http://aladin.u-strasbg.fr/AladinLite/}} developed by CDS\footnote{\url{http://cds.u-strasbg.fr/}}) in the form of a  HIPs color-coded image, as well as each individual tile using \code{VisiOmatic}\citep{2015A&C....10...43B}. The native functions, such as zoom, have been augmented with: 1) an image layer that allows the display of other surveys; 2) a position locator; 3) the possibility of sharing a display with other users; 4) a map viewer that allows the display of \code{HEALpix} maps generated during the processing of the data (\eg, a map of  the number of images available at each point in the sky); 5) different grids and polylines including the original coadd tiles and the border of the DES footprint; 6) access to the image of a specific tile, which can be examined using \code{VisiOmatic}.

The image viewer can be used to examine the tile as a whole and  to inspect a specific position on the sky in more detail using the native functions of the tool that include snapshot, profile overlays, contrast settings, color mix, zoom,  full screen mode, and catalog overlay. The interface also allows side-by-side comparison of the same sky region using different settings. Other useful functions are the ability to switch on/off markers, recenter the display to provide a visualization of the entire tile, crop part of the image, and download the fits images and catalog associated with the tile being displayed.

Finally, one can also overlay object catalogs, which are classified into three categories: 1) targets - list of positions or objects either uploaded or created using the Target viewer and User Query services described below; 2) object catalog - the DES DR1 object catalog produced by DESDM; 3) external catalogs - a sample of catalogs available in Vizier. To facilitate the comparison between catalogs, the user can change the symbol, color, and size of each catalog it selects to display.

\item Target Viewer: This service enables the user to examine a list of uploaded positions of objects selected using the User Query service described below. The first page provides a summary of the available target lists, the ability to upload/delete lists, and the ability to mark favorite lists. Adding a list can be done by pasting a list of (RA, Dec) coordinates.  Once a list is selected, the image surrounding an object/position can be visualized by selecting an entry in the list and the corresponding position. The user can then select the columns to be shown, sort according to a given attribute, and  comment, rank and reject an entry. One can also apply a filter and the filtered list can be downloaded (as \code{csv} or \code{FITS} file)  or saved for future use.

\item User Query: This service provides access to the database table storing the DES DR1 data from which SQL queries can be written, validated, and executed.  The resulting table is displayed under ``My Tables'' where it can be renamed, a few lines of its content listed, and deleted. Objects selected can be immediately viewed in the Target Viewer after the columns of the resulting table are properly associated with those recognized by the tool. 

\end{itemize}

Given the suite of functionalities available in each tool, tutorials in the form of videos have been prepared to help introduce first-time users to the services. Information about the current limitations is available in the help associated with each tool.

\subsection{NOAO Data Lab}

The NOAO Data Lab \citep{2016SPIE.9913E..0LF} is one of the access portals for DES DR1. The goal of the Data Lab is to enable efficient exploration and analysis of large datasets, including catalogs, with a particular focus on surveys using DECam and the NOAO Mosaic cameras. Among its features are a database for catalog data, accessible both from a Table Access Protocol (TAP\footnote{\url{http://ivoa.net/documents/TAP}}) service and from direct PostgreSQL queries, web-based, command-line and programmatic catalog and image query interfaces (both custom and via standard VO protocols), remote storage space for personal database  and files, a JupyterHub-based notebook analysis environment, and a Simple Image Access (SIA) service. The Data Lab hosts the DES DR1 catalog tables using a PostgreSQL v9.6 database on the backend. The data are identical to those hosted by the DES Collaboration through the NCSA portal, but with a few additions; first, the Data Lab database contains tables with crossmatches to other large catalogs (AllWISE, GALEX, HSC, and Simbad) and a table of neighboring objects within a 30 arcsecond radius for all objects in DES. Second, the main table contains a few extra columns, such as Ecliptic and Galactic coordinates, an HTM\footnote{\url{http://www.skyserver.org/htm}} (Hierarchical Triangular Mesh) index, two supplementary \healpix indices in RING (\nside = 256) and NESTED (\nside = 4096) schemes, and precomputed colors. Finally, the data are clustered and indexed using the Q3C scheme \citep{2006ASPC..351..735K}, allowing Q3C functions to perform fast spatial queries on the tables.

\subsubsection{Data Lab Services}
\begin{itemize}
\item Both anonymous and authenticated access to DES DR1 through Data Lab services. Anonymous users of the Data Lab can query the DES DR1 database and use the web-based tools. By creating and logging into an account through \url{datalab.noao.edu}, authenticated users get access to a dedicated Jupyter notebook server with permanently stored notebooks and 1 TB of storage space for personal database tables and files.

\item Access to DES DR1 through TAP. The Data Lab exposes the DES DR1 catalog through a TAP service, which may be accessed through a web query form and schema browser. Alternately, users may query DES DR1 through TOPCAT \citep{2005ASPC..347...29T} by pointing to the URL of the Data Lab TAP service (\url{http://datalab.noao.edu/tap}) from within the application.

\item Python and command-line query clients. The Data Lab client package (\url{https://github.com/noao-datalab/datalab-client}) contains the multipurpose {\tt datalab} command-line interface and the \textit{queryClient} Python module. Both interfaces allow synchronous queries from a personal computer, for which control is suspended until a result is returned, using either ADQL (the native TAP interface query language) or the Postgres SQL syntax. The user may also run asynchronous ADQL queries, for which the query operation is given a job ID and run in the background, through the TAP service. The queryClient module is preinstalled on the Data Lab Jupyter notebook server.

\item An image cutout service. The Data Lab SIA service provides access to cutouts of the DES DR1 images. For a given position on the sky, the SIA service returns a table of metadata of all images that fall within the specified radius. The metadata includes select header information for each image, along with a URL to retrieve a cutout of a specified size.

\item A JupyterHub notebook server and compute environment. This server provides access to common Python libraries as well as all Data Lab Python modules, including the \textit{authClient} authorization module, the \textit{queryClient} query module, the \textit{storeClient} virtual storage module, other interface modules and multiple examples. The Jupyter notebook server provides a convenient way to run code close to the data.

\end{itemize}

More information on using these services is available on the Data Lab web page: \url{http://datalab.noao.edu}. 


\section{Summary and Future Releases}\label{sec:summary}

DES has provided a deep view of the south Galactic cap with precise $grizY$ photometry that will ultimately reach $\roughly24$th magnitude in the $i$ band over $\roughly5000$~deg$^{2}$.
We have described here an overview of the survey, the data acquisition and processing pipelines and gave more details on the release products, the data validation, the known issues and the data access services for the first major public release (DR1).

This release is composed of the reduced images and wide-field coadd source catalogs from the first three years of full science operations, consisting in almost 39,000 single exposure images, close to 62,000 coadd images (including all bands and detection images) covering 10,338 tiles over the DES footprint, resulting in nearly 400M distinct cataloged objects. Benchmark galaxy and stellar samples contain $\roughly310{\rm M}$ and $\roughly80{\rm M}$ objects, respectively, following a basic object quality selection.

The primary attributes of DES DR1 are summarized in \tabref{summary}, and the data products can be accessed from several complementary platforms hosted at NCSA, NOAO, and LIneA, available at \url{https://des.ncsa.illinois.edu/releases/dr1}.
The overall high quality and homogeneity of the release will provide a rich legacy for the international astrophysics community \citep[e.g.,][]{2016MNRAS.460.1270D}.

DES \NEW{finishes its scheduled observations in early 2019} and we expect the next major public DES data release (DR2) will be based on the products available after the survey is completed.


\acknowledgements

Funding for the DES Projects has been provided by the U.S. Department of Energy, the U.S. National Science Foundation, the Ministry of Science and Education of Spain, the Science and Technology Facilities Council of the United Kingdom, the Higher Education Funding Council for England, the National Center for Supercomputing Applications at the University of Illinois at Urbana-Champaign, the Kavli Institute of Cosmological Physics at the University of Chicago, the Center for Cosmology and Astro-Particle Physics at the Ohio State University,
the Mitchell Institute for Fundamental Physics and Astronomy at Texas A\&M University, Financiadora de Estudos e Projetos, Funda{\c c}{\~a}o Carlos Chagas Filho de Amparo {\`a} Pesquisa do Estado do Rio de Janeiro, Conselho Nacional de Desenvolvimento Cient{\'i}fico e Tecnol{\'o}gico and the Minist{\'e}rio da Ci{\^e}ncia, Tecnologia e Inova{\c c}{\~a}o, the Deutsche Forschungsgemeinschaft and the Collaborating Institutions in the Dark Energy Survey. 

The Collaborating Institutions are Argonne National Laboratory, the University of California at Santa Cruz, the University of Cambridge, Centro de Investigaciones Energ{\'e}ticas, Medioambientales y Tecnol{\'o}gicas-Madrid, the University of Chicago, University College London, the DES-Brazil Consortium, the University of Edinburgh, the Eidgen{\"o}ssische Technische Hochschule (ETH) Z{\"u}rich, Fermi National Accelerator Laboratory, the University of Illinois at Urbana-Champaign, the Institut de Ci{\`e}ncies de l'Espai (IEEC/CSIC), the Institut de F{\'i}sica d'Altes Energies, Lawrence Berkeley National Laboratory, the Ludwig-Maximilians Universit{\"a}t M{\"u}nchen and the associated Excellence Cluster Universe, the University of Michigan, the National Optical Astronomy Observatory, the University of Nottingham, The Ohio State University, the University of Pennsylvania, the University of Portsmouth, SLAC National Accelerator Laboratory, Stanford University, the University of Sussex, Texas A\&M University, and the OzDES Membership Consortium.

Based in part on observations at Cerro Tololo Inter-American Observatory, National Optical Astronomy Observatory, which is operated by the Association of Universities for Research in Astronomy (AURA) under a cooperative agreement with the National Science Foundation.

The DES data management system is supported by the National Science Foundation under Grant Numbers AST-1138766 and AST-1536171.
The DES participants from Spanish institutions are partially supported by MINECO under grants AYA2015-71825, ESP2015-66861, FPA2015-68048, SEV-2016-0588, SEV-2016-0597, and MDM-2015-0509, some of which include ERDF funds from the European Union. IFAE is partially funded by the CERCA program of the Generalitat de Catalunya.Research leading to these results has received funding from the European Research Council under the European Union's Seventh Framework Program (FP7/2007-2013) including ERC grant agreements 240672, 291329, and 306478.

Members of LIneA would like to acknowledge the financial support of the special program INCT do  e-Universo

This work made use of the Illinois Campus Cluster, a computing resource
that is operated by the Illinois Campus Cluster Program (ICCP) in conjunction
with the National Center for Supercomputing Applications (NCSA) and which is
supported by funds from the University of Illinois at Urbana-Champaign.

This research is part of the Blue Waters sustained-petascale computing project,
which is supported by the National Science Foundation (awards OCI-0725070 and
ACI-1238993) and the state of Illinois. Blue Waters is a joint effort of the
University of Illinois at Urbana-Champaign and its National Center for
Supercomputing Applications.

We  acknowledge support from the Australian Research Council Centre of Excellence for All-sky Astrophysics (CAASTRO), through project number CE110001020, and the Brazilian Instituto Nacional de Ci\^encia
e Tecnologia (INCT) e-Universe (CNPq grant 465376/2014-2).

This manuscript has been authored by Fermi Research Alliance, LLC under Contract No. DE-AC02-07CH11359 with the U.S. Department of Energy, Office of Science, Office of High Energy Physics. The United States Government retains and the publisher, by accepting the article for publication, acknowledges that the United States Government retains a non-exclusive, paid-up, irrevocable, world-wide license to publish or reproduce the published form of this manuscript, or allow others to do so, for United States Government purposes

\facility{Blanco (DECam)} 
\software{\SExtractor \citep{1996A&AS..117..393B}, \code{PSFEx} \citep{2011ASPC..442..435B}, \code{scamp} \citep{2006ASPC..351..112B}, \code{swarp} \citep{2010ascl.soft10068B}, \mangle \citep{2004MNRAS.349..115H,2008MNRAS.387.1391S}, \healpix \citep{HealpixSoft}\footnote{\url{http://healpix.sourceforge.net}}, \code{astropy} \citep{Astropy:2013}, \code{matplotlib} \citep{2007CSE.....9...90H}, \code{numpy} \citep{numpy:2011}, \code{scipy} \citep{scipy:2001}, \code{healpy}\footnote{\url{https://github.com/healpy/healpy}}, \code{fitsio}\footnote{\url{https://github.com/esheldon/fitsio}}, \code{ngmix} \citep{2014MNRAS.444L..25S}\footnote{\url{https://github.com/esheldon/ngmix}}}, \code{easyaccess}, \code{skymap}

\appendix
\numberwithin{figure}{section}
\numberwithin{table}{section}

\renewcommand{\FIXME}[1]{{\bf \textcolor{red}{#1}}}
\renewcommand{\CHECK}[1]{{\textcolor{orange}{#1}}}
\renewcommand{\COMMENT}[3]{\textcolor{#2}{(#1: #3)}\xspace}

\section{DES Terminology}
\label{app:terminology}


\begin{itemize}

\item \textbf{Image}: a single raw output file corresponding to 1 of the 62 science CCDs in the DECam focal plane.

\item \textbf{Hex}: a hexagonal field on the sky nominally covered by the 62 science CCDs of a single DECam pointing.

\item \textbf{Exposure}: a collection of 62 science CCD images corresponding to a single pointing of DECam.

\item \textbf{Tiling}: a collection of DECam pointings that covers the DES footprint with minimal gaps and overlaps in a single filter.

\item \textbf{Tile}: a sky area unit used by DESDM to parcel the DES footprint and organize the coadd outputs. Each tile is 0\fdg7306 on a side.


\item \textbf{Single-epoch}: relates to the collection and analysis of individual exposures for a single band prior to coaddition.

\item \textbf{Coadd}: the process of combining the data from multiple exposures over an area in order to increase depth.

\item \textbf{Y3A2 Release}: second annual internal release by DESDM to the DES Collaboration of data products obtained from the first three seasons of DES science operations and the Science Verification period. DR1 is based on Y3A2.

\end{itemize}

\section{Photometric Transformations}
\label{app:transform}

We provide empirical photometric transformations between DES and other surveys that were used for the validation of DR1 data products.





\subsection{Gaia}

A sample of stars taken from the deep supernova fields with at least 25 observations in each of the $g$, $r$, and $i$ bands were matched to Gaia stars with $G < 20 
\magn$. The Gaia $G$-band magnitudes predicted from from DES photometry are

\begin{equation}
G = r - 0.100 + 0.150 (g-i) - 0.013 (g-i)^2 - 0.035 (g-i)^3,
\end{equation}

\noindent valid for stars with $0.3< g-i <3.0$.


\subsection{CFHTLenS}

As described in \secref{depth}, we transformed photometry from the CFHTLenS survey to the DES system to report the detection efficiency of CFHTLenS objects as a function of magnitude in the DES system. 
The corresponding transformation equations, which were empirically derived from matched stars with high measured signal-to-noise are:

\begin{equation}
\begin{aligned}
g_{\rm DES} = g_{\rm CFHTLenS}  + 0.040 (g - r)_{\rm CFHTLenS} + 0.143 \\
r_{\rm DES} = r_{\rm CFHTLenS}  - 0.083 (g - r)_{\rm CFHTLenS} + 0.089 \\
i_{\rm DES} = i_{\rm CFHTLenS}  - 0.179 (i - z)_{\rm CFHTLenS} + 0.132 \\
z_{\rm DES} = z_{\rm CFHTLenS}  - 0.067 (i - z)_{\rm CFHTLenS} + 0.115
\end{aligned}
\end{equation}



\section{Example queries}\label{app:queries}

We provide several example SQL queries to the DR1 Oracle database tables. Note that these queries can be easily modified to ADQL language used by the TAP service at NOAO Data Labs.





\begin{enumerate}

\item \textbf{Sample of objects:} This query returns a sample of 0.0001\% of objects from DR1.

\footnotesize
\begin{lstlisting}[language=sql]
SELECT ra, dec, mag_auto_g FROM dr1_main SAMPLE(0.0001);
\end{lstlisting}
\normalsize

Also, one can select the first rows of a table with a command like

\footnotesize
\begin{lstlisting}[language=sql]
SELECT ra, dec, mag_auto_g FROM dr1_main WHERE ROWNUM <= 10;
\end{lstlisting}
\normalsize

\item \textbf{Select stars from the M2 globular cluster:} This query uses the \var{extended\_coadd} example morphological classifier to separate point-like stars from spatially extended galaxies (see \secref{sgsep}). An interstellar reddening correction has been pre-applied to the columns marked with the \var{\_dered} suffix (see \secref{photometry}). This query was used to generate \figref{globular_cluster}.

\footnotesize
\begin{lstlisting}[language=sql]
SELECT
  	coadd_object_id,
	ra, dec, ebv_sfd98,
	mag_auto_g_dered,
	mag_auto_r_dered,
	wavg_mag_psf_g_dered,
	wavg_mag_psf_r_dered
FROM
	dr1_main
WHERE
   	ra BETWEEN 323.36 - 0.125 AND 323.36 + 0.125
   	AND dec BETWEEN -0.82 - 0.125 AND -0.82 + 0.125
	AND ( (CASE WHEN spread_model_i + 3. * spreaderr_model_i > 0.005 THEN 1 ELSE 0 END) +
  	      (CASE WHEN spread_model_i + 1. * spreaderr_model_i > 0.003 THEN 1 ELSE 0 END) +
  	      (CASE WHEN spread_model_i - 1. * spreaderr_model_i > 0.003 THEN 1 ELSE 0 END) ) <= 1
   	AND spread_model_i BETWEEN -0.05 AND 0.05
   	AND imaflags_iso_g = 0
   	AND imaflags_iso_r = 0
   	AND flags_g < 4
	AND flags_r < 4;
\end{lstlisting}
\normalsize

\item \textbf{Create galaxy density map:} Create a \healpix (NEST schema, celestial coordinates) galaxy density map at resolution \nside = 1024 (\roughly3.4\arcmin)


\footnotesize
\begin{lstlisting}[language=sql]
SELECT
	hpix_1024, COUNT(mag_auto_i), AVG(mag_auto_i)
FROM
	dr1_main
WHERE
	( (CASE WHEN spread_model_i + 3. * spreaderr_model_i > 0.005 THEN 1 ELSE 0 END) +
  	  (CASE WHEN spread_model_i + 1. * spreaderr_model_i > 0.003 THEN 1 ELSE 0 END) +
  	  (CASE WHEN spread_model_i - 1. * spreaderr_model_i > 0.003 THEN 1 ELSE 0 END) ) = 3
	AND spread_model_i BETWEEN -0.05 AND 0.05
	AND imaflags_iso_i = 0
	AND flags_i < 4
	AND mag_auto_i < 23
GROUP BY
	hpix_1024;
\end{lstlisting}
\normalsize

\item \textbf{Return urls for the complete coadd object catalogs and $grizY$ coadd images within a small patch of sky:}

\footnotesize
\begin{lstlisting}[language=sql]
SELECT
	tilename, 
	fits_dr1_main, 
	fits_image_g, fits_image_r, fits_image_i, fits_image_z, fits_image_y  
FROM
	dr1_tile_info
WHERE
	ra_cent BETWEEN 30. AND 32.
    	AND dec_cent BETWEEN -6. AND -4.;
\end{lstlisting}
\normalsize


\end{enumerate}

\section{SExtractor Flags}
\label{app:sextractor_flags}

\NEW{\tabref{sextractor_flags} summarizes the standard warning flags provided by \sextractor, encoded in the \var{FLAGS} bitmask column for each band.}

\begin{deluxetable}{l c}[!hb]
\centering
\tablewidth{0pt}
\tabletypesize{\scriptsize}
\tablecaption{Summary of bitmask values and warning descriptions for the \sextractor \var{FLAGS} column\tablenotemark{a} \label{tab:sextractor_flags}}
\tablehead{
  \colhead{Bit} & \colhead{Description}
}
\startdata
1 & The object has neighbours, bright and close enough to significantly bias the \\
 & MAG AUTO photometry, or bad pixels (more than 10\% of the integrated area affected) \\
2 & The object was originally blended with another one \\
4 & At least one pixel of the object is saturated (or very close to) \\
8 & The object is truncated (too close to an image boundary) \\
16 & Object's aperture data are incomplete or corrupted \\
32 & Object's isophotal data are incomplete or corrupted \\
64 & A memory overflow occurred during deblending \\
128 & A memory overflow occurred during extraction \\
\enddata
\tablenotetext{a}{Table data obtained from \url{https://www.astromatic.net/pubsvn/software/sextractor/trunk/doc/sextractor.pdf}.}
\end{deluxetable}

\newpage
\section{Released Tables}
\label{app:appendix_tables}
The DR1 catalog data is mostly comprised of four tables. \drmain includes all the main quantities extracted from
the coadd pipeline and important information about the objects. That table also includes \magauto and \var{wavg\_mag\_psf}, as well as the corresponding dereddened magnitudes. \drflux and \drmagnitude contain 15 different measurements of the fluxes and magnitudes for each object. Additionally, these three tables share some commonly used columns to facilitate queries (by avoiding the need to join multiple tables). The fourth table, \drtileinfo contains information relevant to the processed tiles, from the tile geometry to the urls of associated files.

%
%
%

\startlongtable
\begin{deluxetable}{l l c}
    	\centering
	\tablewidth{0pt}
	\tabletypesize{\scriptsize}
	\tablecaption{DR1\_MAIN Table description: 399,263,026 rows; 213 columns \label{tab:dr1_main}}
	\tablehead{
		\colhead{Column Name} & \colhead{Description} & \colhead{Number of columns}
	}
        \startdata
        COADD\_OBJECT\_ID & Unique identifier for the coadded objects & 1 \\
        TILENAME & Identifier of each one of the tiles on which the survey is gridded & 1 \\
        RA & Right ascension, with quantized precision for indexing (ALPHAWIN\_J2000 has full precision  & 1 \\
         & but not indexed) [degrees] &  \\
        ALPHAWIN\_J2000 & Right ascension for the object, J2000 in ICRS system (full precision but not indexed) [degrees] & 1 \\
        DEC & Declination, with quantized precision for indexing (DELTAWIN\_J2000 has full precision but not indexed) [degrees] & 1 \\
        DELTAWIN\_J2000 & Declination for the object, J2000 in ICRS system (full precision but not & 1 \\
         & indexed) [degrees] &  \\
        GALACTIC\_L & Galactic Longitude [degrees] & 1 \\
        GALACTIC\_B & Galactic Latitude [degrees] & 1 \\
        XWIN\_IMAGE & X-centroid from windowed measurements on coadded image [pixel] & 1 \\
        YWIN\_IMAGE & Y-centroid from windowed measurements on coadded image [pixel] & 1 \\
        XWIN\_IMAGE\_{G,R,I,Z,Y} & X-centroid from windowed measurements on coadded band image [pixel] & 5 \\
        YWIN\_IMAGE\_{G,R,I,Z,Y} & Y-centroid from windowed measurements on coadded band images [pixel] & 5 \\
        X2WIN\_IMAGE\_{G,R,I,Z,Y} & Second moment in x-direction, from converged windowed measurements [pixel2] & 5 \\
        ERRX2WIN\_IMAGE\_{G,R,I,Z,Y} & Uncertainty in second moment of x-distribution centroid, from converged windowed measurements [pixel2] & 5 \\
        Y2WIN\_IMAGE\_{G,R,I,Z,Y} & Second moment in y-direction, from converged windowed measurements [pixel2] & 5 \\
        ERRY2WIN\_IMAGE\_{G,R,I,Z,Y} & Uncertainty in second moment of y-distribution centroid, from converged windowed measurements [pixel2] & 5 \\
        XYWIN\_IMAGE\_{G,R,I,Z,Y} & Second moment in xy-direction, from converged windowed measurements [pixel2] & 5 \\
        ERRXYWIN\_IMAGE\_{G,R,I,Z,Y} & Uncertainty in second moment of xy-distribution, from converged windowed measurements [pixel2] & 5 \\
        HPIX\_{32,64,1024,4096,16384} & Healpix identifier for its \nside grid size, in a NESTED schema & 5 \\
        NEPOCHS\_{G,R,I,Z,Y} & Number of epochs the source is detected in single epoch images & 5 \\
        NITER\_MODEL\_{G,R,I,Z,Y} & Number of iterations in model fitting photometric measurements & 5 \\
        ISOAREA\_IMAGE\_{G,R,I,Z,Y} & Isophotal area of the coadded source [pixel2] & 5 \\
        A\_IMAGE & Major axis size based on an isophotal model [pixel] & 1 \\
        ERRA\_IMAGE & Uncertainty in major axis size, from isophotal model [pixel] & 1 \\
        AWIN\_IMAGE\_{G,R,I,Z,Y} & Major axis size, from 2nd order windowed moment measurements [pixels] & 5 \\
        ERRAWIN\_IMAGE\_{G,R,I,Z,Y} & Uncertainty in major axis size, from converged windowed measurement, assuming uncorrelated noise [pixel] & 5 \\
        B\_IMAGE & Minor axis size based on an isophotal model [pixel] & 1 \\
        ERRB\_IMAGE & Uncertainty in minor axis size, from isophotal model [pixel] & 1 \\
        BWIN\_IMAGE\_{G,R,I,Z,Y} & Minor axis size, from 2nd order windowed moment measurements [pixels] & 5 \\
        ERRBWIN\_IMAGE\_{G,R,I,Z,Y} & Uncertainty in minor axis size, from converged windowed measurement, assuming uncorrelated noise [pixel] & 5 \\
        THETA\_J2000 & Position angle of source in J2000 coordinates, from non-windowed measurement [degrees] & 1 \\
        ERRTHETA\_IMAGE & Uncertainty in source position, from isophotal model [degrees] & 1 \\
        THETAWIN\_IMAGE\_{G,R,I,Z,Y} & Position angle of source, for converged windowed measurement grow from x to y [degrees] & 5 \\
        ERRTHETAWIN\_IMAGE\_{G,R,I,Z,Y} & Uncertainty in source position angle, from converged windowed measurement [degrees] & 5 \\
        FWHM\_IMAGE\_{G,R,I,Z,Y} & FWHM measured from the isophotal area, from elliptical growth-curve, modeled in 2 dimensions [pixel] & 5 \\
        FLUX\_RADIUS\_{G,R,I,Z,Y} & Half-light radius for the object, from elliptical growth-curve, modeled in 2 dimensions [pixel] & 5 \\
        KRON\_RADIUS & Kron radius measured from detection image [pixel] & 1 \\
        KRON\_RADIUS\_{G,R,I,Z,Y} & Kron radius measured from coadded image [pixel] & 5 \\
        CLASS\_STAR\_{G,R,I,Z,Y} & Simple morphological extended source classifier. Values between 0 (galaxies) and 1 (stars). & 5 \\
         & SPREAD\_MODEL exhibits better performance for morphological classification. &  \\
        SPREAD\_MODEL\_{G,R,I,Z,Y} & Morphology based classifier based on comparison between a PSF versus exponential-PSF model. & 5 \\
         & Values closer to 0 correspond to stars, larger values correspond to galaxies &  \\
        SPREADERR\_MODEL\_{G,R,I,Z,Y} & Uncertainty in morphology based classifier based on comparison between PSF versus exponential-PSF model. & 5 \\
        WAVG\_SPREAD\_MODEL\_{G,R,I,Z,Y} & SPREAD MODEL using the weighted averaged values from single epoch detections & 5 \\
        WAVG\_SPREADERR\_MODEL\_{G,R,I,Z,Y} & Uncertainty in SPREAD MODEL using the weighted averaged values from single epoch detections & 5 \\
        FLUX\_AUTO\_{G,R,I,Z,Y} & Aperture-flux measurement, elliptical model based on the Kron radius [ADU] & 5 \\
        FLUXERR\_AUTO\_{G,R,I,Z,Y} & Uncertainty in aperture-flux measurement, elliptical model based on the Kron radius [ADU] & 5 \\
        WAVG\_FLUX\_PSF\_{G,R,I,Z,Y} & Weighted average flux measurement of PSF fit single epoch detections [ADU] & 5 \\
        WAVG\_FLUXERR\_PSF\_{G,R,I,Z,Y} & Uncertainty of weighted averaged flux measurement of PSF fit single epoch detections [ADU] & 5 \\
        MAG\_AUTO\_{G,R,I,Z,Y} & Magnitude estimation, for an elliptical model based on the Kron radius [mag] & 5 \\
        MAGERR\_AUTO\_{G,R,I,Z,Y} & Uncertainty in magnitude estimation, for an elliptical model based on the Kron radius [mag] & 5 \\
        MAG\_AUTO\_{G,R,I,Z,Y}\_DERED & Dereddened magnitude estimation (using SFD98), for an elliptical model based on the Kron radius [mag] & 5 \\
        WAVG\_MAG\_PSF\_{G,R,I,Z,Y} & Weighted average magnitude, of PSF fit single epoch detections [mag] & 5 \\
        WAVG\_MAGERR\_PSF\_{G,R,I,Z,Y} & Uncertainty of weighted averaged magnitude, of PSF fit single epoch detections [mag] & 5 \\
        WAVG\_MAG\_PSF\_{G,R,I,Z,Y}\_DERED & Dereddened weighted average magnitude (using SFD98)  from PSF fit single epoch detections [mag] & 5 \\
        EBV\_SFD98 & E(B-V) reddening coefficient from Schlegel, Finkbeiner \& Davis, 1998 [mag] & 1 \\
        BACKGROUND\_{G,R,I,Z,Y} & Background level, by CCD-amplifier [mag] & 5 \\
        FLAGS\_{G,R,I,Z,Y} & Additive flag describing cautionary advice about source extraction process. Use less than 4 for well behaved objects & 5 \\
        IMAFLAGS\_ISO\_{G,R,I,Z,Y} & Flag identifying sources with missing/flagged pixels, considering all single epoch images & 5 \\
	\enddata
\end{deluxetable}

\clearpage

\startlongtable
\begin{deluxetable}{l l c}
	\centering
	\tablewidth{0pt}
	\tabletypesize{\scriptsize}
    	\tablecaption{DR1\_FLUX Table Description: 399,263,026 rows, 179 columns \label{tab:dr1_flux}}
	\tablehead{
		\colhead{Column Name} & \colhead{Description} & \colhead{Number of columns}
	}
	\startdata
	COADD\_OBJECT\_ID & Unique identifier for the coadded objects & 1 \\
	TILENAME & Identifier of each one of the tiles on which the survey is gridded & 1 \\
	RA & Right ascension, with quantized precision for indexing (ALPHAWIN\_J2000 has full precision but not indexed) [degrees] & 1 \\
	ALPHAWIN\_J2000 & Right ascension for the object, J2000 in ICRS system (full precision but not indexed) [degrees] & 1 \\
	DEC & Declination, with quantized precision for indexing (DELTAWIN\_J2000 has full precision but not indexed) [degrees] & 1 \\
	DELTAWIN\_J2000 & Declination for the object, J2000 in ICRS system (full precision but not indexed) [degrees] & 1 \\
	XWIN\_IMAGE & X-centroid of source (from coadd detection image) [pixel] & 1 \\
	YWIN\_IMAGE & Y-centroid of source (from coadd detection image) [pixel] & 1 \\
	HPIX\_{32,64,1024,4096,16384} & Healpix identifier for its \nside grid size, in a NESTED schema & 5 \\
	FLUX\_AUTO\_{G,R,I,Z,Y} & Flux measurement, for an elliptical model based on the Kron radius [ADU] & 5 \\
	FLUXERR\_AUTO\_{G,R,I,Z,Y} & Uncertainty in flux measurement, for an elliptical model based on the Kron radius [ADU] & 5 \\
	FLUX\_APER\_{1-12}\_{G,R,I,Z,Y} & Flux measurement for circular apertures [ADU] & 60 \\
	FLUXERR\_APER\_{1-12}\_{G,R,I,Z,Y} & Uncertainty in flux measurement for circular apertures [ADU] & 60 \\
	FLUX\_PETRO\_{G,R,I,Z,Y} & Flux for a Petrosian radius [ADU] & 5 \\
	FLUXERR\_PETRO\_{G,R,I,Z,Y} & Uncertainty in flux for a Petrosian radius [ADU] & 5 \\
	WAVG\_FLUX\_PSF\_{G,R,I,Z,Y} & Weighted averaged flux, of PSF fit single epoch detections [ADU] & 5 \\
	WAVG\_FLUXERR\_PSF\_{G,R,I,Z,Y} & Uncertainty of weighted averaged flux, of PSF fit single epoch detections [ADU] & 5 \\
	PETRO\_RADIUS\_{G,R,I,Z,Y} & Petrosian radius [pixel] & 5 \\
	EBV\_SFD98 & E(B-V) reddening coefficient from Schlegel, Finkbeiner \& Davis, 1998 [mag] & 1 \\
	FLAGS\_{G,R,I,Z,Y} & Additive flag describing cautionary advice about source extraction process. Use less than 4 for well behaved objects & 5 \\
	IMAFLAGS\_ISO\_{G,R,I,Z,Y} & Flag identifying sources with missing/flagged pixels, considering all single epoch images & 5 \\
	\enddata
\end{deluxetable}

\clearpage

\startlongtable
\begin{deluxetable}{l l c}
 	\centering
	\tablewidth{0pt}
	\tabletypesize{\scriptsize}
	\tablecaption{DR1\_MAGNITUDE Table Description: 399,263,026 rows, 179 columns \label{tab:dr1_mag}}
	\tablehead{
		\colhead{Column Name} & \colhead{Description} & \colhead{Number of columns}
	}
        \startdata
        COADD\_OBJECT\_ID & Unique identifier for the coadded objects & 1 \\
        TILENAME & Identifier of each one of the tiles on which the survey is gridded & 1 \\
        RA & Right ascension, with quantized precision for indexing (ALPHAWIN\_J2000 has full precision but not indexed) [degrees] & 1 \\
        ALPHAWIN\_J2000 & Right ascension for the object, J2000 in ICRS system (full precision but not indexed) [degrees] & 1 \\
        DEC & Declination, with quantized precision for indexing (DELTAWIN\_J2000 has full precision but not indexed) [degrees] & 1 \\
        DELTAWIN\_J2000 & Declination for the object, J2000 in ICRS system (full precision but not indexed) [degrees] & 1 \\
        XWIN\_IMAGE & X-centroid from windowed measurements on coadded image [pixel] & 1 \\
        YWIN\_IMAGE & Y-centroid from windowed measurements on coadded image [pixel] & 1 \\
        HPIX\_{32,64,1024,4096,16384} & Healpix identifier for its \nside grid size, in a NESTED schema & 5 \\
        MAG\_AUTO\_{G,R,I,Z,Y} & Magnitude estimation, for an elliptical model based on the Kron radius [mag] & 5 \\
        MAGERR\_AUTO\_{G,R,I,Z,Y} & Uncertainty in magnitude estimation, for an elliptical model based on the Kron radius [mag] & 5 \\
        MAG\_APER\_{1-12}\_{G,R,I,Z,Y} & Magnitude estimation for circular apertures [mag] & 60 \\
        MAGERR\_APER\_{1-12}\_{G,R,I,Z,Y} & Uncertainty in magnitude estimation for circular apertures [mag] & 60 \\
        MAG\_PETRO\_{G,R,I,Z,Y} & Magnitude for a Petrosian radius [mag] & 5 \\
        MAGERR\_PETRO\_{G,R,I,Z,Y} & Uncertainty in magnitude for a Petrosian radius [mag] & 5 \\
        WAVG\_MAG\_PSF\_{G,R,I,Z,Y} & Weighted average magnitude, of PSF fit single epoch detections [mag] & 5 \\
        WAVG\_MAGERR\_PSF\_{G,R,I,Z,Y} & Uncertainty of weighted averaged magnitude, of PSF fit single epoch detections [mag] & 5 \\
        PETRO\_RADIUS\_{G,R,I,Z,Y} & Petrosian radius [pixel] & 5 \\
        EBV\_SFD98 & E(B-V) reddening coefficient from Schlegel, Finkbeiner \& Davis, 1998 [mag] & 1 \\
        FLAGS\_{G,R,I,Z,Y} & Additive flag describing cautionary advice about source extraction process. Use less than 4 for well behaved objects & 5 \\
        IMAFLAGS\_ISO\_{G,R,I,Z,Y} & Flag identifying sources with missing/flagged pixels, considering all single epoch images & 5 \\
        \enddata
\end{deluxetable}

\clearpage

\startlongtable
\begin{deluxetable}{l l c}
	\centering
	\tablewidth{0pt}
	\tabletypesize{\scriptsize}
	\tablecaption{DR1\_TILE\_INFO Table Description: 10,338 rows, 46 columns \label{tab:dr1_tile}}
	\tablehead{
		\colhead{Column Name} & \colhead{Description} & \colhead{Number of columns}
	}
        \startdata
        TILENAME & Tilename identifier & 1 \\
        RA\_CENT & Central Right Ascension for tile [degrees] & 1 \\
        DEC\_CENT & Central Declination for tile [degrees] & 1 \\
        COUNT & Number of objects per tile & 1 \\
        RAC1 & Right Ascension at Corner 1 of tile [degrees] & 1 \\
        RAC2 & Right Ascension at Corner 2 of tile [degrees] & 1 \\
        RAC3 & Right Ascension at Corner 3 of tile [degrees] & 1 \\
        RAC4 & Right Ascension at Corner 4 of tile [degrees] & 1 \\
        RACMAX & Maximum Right Ascension covered in tile [degrees] & 1 \\
        RACMIN & Minimum Right Ascension covered in tile [degrees] & 1 \\
        RA\_SIZE & Extent of Right Ascension for tile [degrees] & 1 \\
        URAMAX & Maximum Unique Right Ascension of objects measured from tile [degrees] & 1 \\
        URAMIN & Minimum Unique Right Ascension of objects measured from tile [degrees] & 1 \\
        DECC1 & Declination at Corner 1 of tile [degrees] & 1 \\
        DECC2 & Declination at Corner 2 of tile [degrees] & 1 \\
        DECC3 & Declination at Corner 3 of tile [degrees] & 1 \\
        DECC4 & Declination at Corner 4 of tile [degrees] & 1 \\
        DECCMAX & Maximum Declination covered in tile [degrees] & 1 \\
        DECCMIN & Minimum Declination covered in tile [degrees] & 1 \\
        DEC\_SIZE & Extent of Declination for tile, in average is 0.7304 deg [degrees] & 1 \\
        UDECMAX & Maximum Unique Declination of objects measured from tile [degrees] & 1 \\
        UDECMIN & Minimum Unique Declination of objects measured from tile [degrees] & 1 \\
        CTYPE1 & WCS projection used for axis 1. Value: RA---TAN & 1 \\
        CTYPE2 & WCS projection used for axis 2. Value: DEC--TAN & 1 \\
        NAXIS1 & WCS definition for number of pixels for axis 1 & 1 \\
        NAXIS2 & WCS definition for number of pixels for axis 2 & 1 \\
        CRPIX1 & WCS definition of central pixel for axis 1. Value: 5000.5 & 1 \\
        CRPIX2 & WCS definition of central pixel for axis 2. Value: 5000.5 & 1 \\
        CRVAL1 & WCS definition of central pixel value for axis 1 & 1 \\
        CRVAL2 & WCS definition of central pixel value for axis 2 & 1 \\
        CD1\_1 & WCS definition for pixel orientation. Value: -0.0000730556 & 1 \\
        CD1\_2 & WCS definition for pixel orientation. Value: 0 & 1 \\
        CD2\_1 & WCS definition for pixel orientation. Value: 0 & 1 \\
        CD2\_2 & WCS definition for pixel orientation. Value: 0.0000730556 & 1 \\
        CROSSRA0 & Flag tile bounday crosses RA=0/24h boundary. Values: Y or N & 1 \\
        PIXELSCALE & WCS definition of pixel scale. Values: 0.263 arcsec/pixel [arcsec/pixel] & 1 \\
        TIFF\_COLOR\_IMAGE & Filename of the TIFF image for the tile, being {tilename}\_r\{reqnum\}p\{attnum\}\_irg.tiff & 1 \\
        FITS\_DR1\_FLUX & Filename of the served FITS being {tilename}\_dr1\_flux.fits.fz & 1 \\
        FITS\_DR1\_MAGNITUDE & Filename of the served FITS being {tilename}\_dr1\_magnitude.fits.fz & 1 \\
        FITS\_DR1\_MAIN & Filename of the served FITS being {tilename}\_dr1\_main.fits.fz & 1 \\
        FITS\_IMAGE\_DET & Filename of the served FITS being {tilename}\_r\{reqnum\}p\{attnum\}\_det.fits.fz & 1 \\
        FITS\_IMAGE\_{G,R,I,Z,Y} & Filename of the served FITS, per band, being {tilename}\_r\{reqnum\}p\{attnum\}\_{band}.fits.fz & 5 \\
        \enddata
\end{deluxetable}

\clearpage


\bibliographystyle{aasjournal}
\bibliography{biblio}

\end{document}